\documentclass[11pt]{article} 

\usepackage[numbers]{natbib} 
\usepackage[affil-it]{authblk}  

\usepackage{geometry} 
\usepackage{graphicx} 
\usepackage{breqn} 
\usepackage{caption} 

\usepackage{booktabs} 
\usepackage{array} 
\usepackage{paralist} 
\usepackage{verbatim} 
\usepackage{subfig} 
\usepackage{amsmath} 
\usepackage{csvsimple} 

\usepackage{fancyhdr} 
\usepackage{sectsty}
\allsectionsfont{\sffamily\mdseries\upshape} 

\usepackage[nottoc,notlof,notlot]{tocbibind} 
\usepackage[titles,subfigure]{tocloft} 

\usepackage[title, page]{appendix}

\title{`Dark pressure': A metric that yields distance-independent circular orbital speed in deep space with `tunable' total mass.}
\author[*]{C. Mallary}
\date{} 

\begin{document}
\maketitle

\begin{abstract}
In general relativity, the tangential speed of objects in stable circular orbits is not uniquely described by the orbital radius and the mass present inside the orbital radius. This work presents a static, spherically symmetric spacetime metric which produces stable circular orbits whose speed approaches a constant mass-independent value at high radii, potentially reducing the need for dark matter to explain the orbital speed of objects far from a galactic center. However, the metric does not contain rotational components,  and therefore requires pressure throughout the spacetime to keep it static. The stress energy tensor of this metric is evaluated numerically using the mass of the Milky Way's central black hole, the orbital speed of its distant satellites, and three different values of a unitless `mass tuning' parameter $B$. These $B$  `tune' the amount of mass present, without violating the Weak Energy Condition (WEC) at any evaluated spacetime point. The stress energy tensor is a solution to the Generalized Tolman-Oppenheimer-Volkoff (GTOV) equation for anisotropic hydrostatic equilibrium, indicating that the metric solves the Einstein Field Equations for an imperfect fluid with a particular density and pressure profile. The metric can be merged with a Friedmann-Robertson-Walker metric, in which case it achieves isotropy at cosmological distances.
\end{abstract}

\section{Introduction}\label{sec:intro}

 Objects far out in the galactic disc orbit the center at a tangential speed that appears to be independent of distance from the center. This is generally attributed to the galaxy having a mass much greater than its visible mass, i.e. that galaxies contain large amounts of dark matter. This conclusion is sensible under Newtonian dynamics. However, general relativity (GR) dynamics are far more complicated than Newtonian dynamics. Observed galactic rotation curves are the primary reason for hypothesizing dark matter in galaxies like ours. The amount of undetected matter needed to maintain galactic rotation curves may be significantly less under GR than under Newtonian approximations.

Cooperstock and Tieu \cite{cooperstock} described a general relativistic galactic model that appeared to account for galactic rotation curves using only luminous and non-exotic dark matter. In their model, the galaxy consists of a rotating disc of pressure-free dust. The authors note that the rotation of the disc serves to keep the model static in the absence of pressure. However, a rotating system is significantly more complicated than a non-rotating one, and analysis is correspondingly difficult; later works argued that the Coopersock \& Tieu model failed to represent rotation curves without generating exotic matter \cite{vogt2005}. 

More recently, Crosta, et al. \cite{Crosta_2020} \cite{beordo_crosta} described a model based on work by Balasin and Grumiller (BG) \cite{BALASIN_2008}. Crosta, et al. purport to use a rotating, pressureless dust to generate Milky Way rotation curves. Although this model is interesting, it fails to represent the region above and below the galactic bulge. The model simply `cuts out' this region of spacetime, instead of patching it to a metric that is accurate in the cutout region. This cutout is acceptable in the orginal BG work, where the model is explicitly presented as a toy (Section 3.2 of \cite{BALASIN_2008}), but it is a weakness if this model is intended to be realistic: Although the model can be made to fit observed orbital data in the galactic disc, it cannot be made to fit any known galactic-scale shape of spacetime. Moreover, the Crosta model cannot claim to be pressureless, since the Einstein equations presented there and in \cite{BALASIN_2008} depend on derivatives of a function $\nu(r,z)$, which is never given analytic form by Crosta, and which Crosta, et al, explicitly approximate as constant. This approximation produces significant pressure in the spacetime (Appendix \ref{app:crosta-pressure}). The BG/Crosta model has also experienced challenges based on its reference frames \cite{Costa_2023}. Ultimately, although Crosta's model attempts to show that the Milky Way does not need dark matter to sustain its rotation curves, it is too incomplete to demonstrate its underlying principle: Namely, that orbits need not depend on mass alone. 

This work takes a different approach: We start with a simplified galactic rotation curve, and find the simplest physically reasonable metric which can produce it without relying exclusively on mass.  The result is a non-rotating, spherically symmetric `galaxy' metric, where pressure keeps the spacetime static, instead of rotation. The pressure also supports circular orbits whose speed is asymptotically constant at high orbital radius. We call this a `dark pressure' metric, because the pressure serves a similar function to dark matter in other models. The dark pressure metric is explicitly a toy: `Dark pressure' does not correspond to any known phenomena of the Milky Way. However, our model has the virtue of simplicity:  The entire spacetime is covered by a single, continuous, diagonal analytic metric, removing the possibility that nonphysical phenomena could arise at cutouts or metric junctions. This simple metric can be numerically tuned to represent a wide range of galactic masses without affecting the speed of distant circular orbits. This decoupling of mass and orbital speed has no Newtonian analog. It is not frame-dragging, though it may be conceptually related:  In GR, orbits do not depend solely on mass. 

The dark pressure metric is not Modified Newtonian Dynamics (MoND) \cite{MOND}. It requires no modification to either GR or Newton's laws. Fundamentally, the dark pressure metric is a demonstration that general relativity can produce counterintuitive results for mass and gravitation.  

This paper follows the structure below:

\begin{itemize}

\item Sec.~\ref{sec:methods}:  Methods

	\begin{itemize}
		\item Sec. \ref{sec:math-basis}:  The mathematical basis for mass-independent orbits in GR.
	
		\item Sec. \ref{sec:gtt}:  Finding a form for the GR metric component $g_{tt}$ which produces constant-speed circular orbits at high radial distance $r$.
	
		\item Sec. \ref{sec:grr}:  Proposing a form for $g_{rr}$. 
	
	\item Sec \ref{sec:FRW-mod}:  Merging the `Basic' metric found in Secs.~\ref{sec:gtt},\ref{sec:grr} with a cosmological metric. This is necesssary for realism at cosmological $r$ because the Basic metric is not perfectly asymptotically flat, so the metric must be modified to be dominated by cosmological features at cosmological scales. We globally alter the Basic metric so that it is asymptotically an FRW metric. 

	\end{itemize}

\item Sec.~\ref{sec:results}:  Results
	\begin{itemize}
		\item Sec.~\ref{sec:analytic-basic}: Analytic Weak Energy Condition compliance of the Basic metric.
		\item Sec.~\ref{sec:inputs}: Input parameters for numeric analysis of the Basic and FRW-modified `dark pressure' metric, based on Milky Way and cosmological data.
		\item Sec.~\ref{sec:numeric} Numeric outputs of the stress-energy tensor $T^{\hat{\mu}\hat{\nu}}$, and their compliance with the Weak Energy Condition (WEC) and Friedmann equations.
		\item Sec.~\ref{sec:particle-density} Particle density of a `dark pressure' galaxy, comparison to Milky Way.
		\item Sec.~\ref{sec:mass-contained} Mass contained within a `dark pressure' galaxy, comparison to Milky Way.
		\item Sec.~\ref{sec:gtov}  The Basic (non-expanding) metric's analytic compliance with the Generalized Tolman-Oppenheimer-Volkoff (GTOV) equation for anisotropic hydrostatic equilibrium.

	\end{itemize}

\item Sec.~\ref{sec:discussion}: Discussion of the results and conclusion.

\item Sec.~\ref{sec:acknowledgements}: Acknowledgements.
\end{itemize}

Unit conversions are discussed in Appendix \ref{app:units}.

\section{Methods}\label{sec:methods}

\subsection{Mathematical basis for mass-independent circular orbits.}\label{sec:math-basis}

Consider a diagonal spacetime metric $g_{\mu\nu}$ in spherical coordinates $(t,r,\theta,\phi)$, which depends only on radius $r$ and has no curvature in the angular dimensions. Written in line-element form, this metric is:

\begin{eqnarray}\label{eq:generic-metric}
ds^2 &=& -|g_{tt}(r)| dt^2 + g_{rr}(r) dr^2 + r^2 d\theta^2 + r^2 \sin^2{\theta}d\phi^2 \nonumber \\
&=&  -|g_{tt}(r)| dt^2 + g_{rr}(r) dr^2 + g_{\Omega\Omega}d\Omega^2
\end{eqnarray}

Here, $g_{\Omega\Omega}$ is used as shorthand for the angular components of the metric.

Evaluating the geodesics for Eq.~\ref{eq:generic-metric} at $\theta = \pi/2$ (the equator), we will see that the speed of a stable circular orbit depends only on $g_{tt}$ and $g_{\Omega\Omega}$, and $g_{rr}$ is irrelevant (see Eq.~\ref{eq:this-must-be-true} for results on a Schwarzchild-like metric).  Evaluating the stress-energy tensor for this metric, we will see that the mass-energy density represented by this metric depends only on $g_{rr}$, and $g_{tt}$ and $g_{\Omega\Omega}$ are irrelevant (see Eq.~\ref{eq:rho-basic-generic}).  Therefore, mass and orbital speed are not directly coupled in GR: We cannot assume we know the mass density of the galaxy based on the speed of stable circular orbits.

In practice, $g_{tt}$ and $g_{rr}$ are constrained by the need to have energetically realistic conditions in the spacetime.  Namely, no timelike observer in this spacetime should see negative mass-energy density anywhere.  This is the Weak Energy Condition (WEC)\cite{Curiel2017}. 

Equivalently, WEC says that, for mass-energy density $\rho = T^{\hat{0}\hat{0}}$ and every directional pressure $P_i = T^{\hat{i}\hat{i}}$:
\begin{eqnarray}\label{eq:wec}
\rho \geq 0 \nonumber \\ 
\rho + P_i \geq 0
\end{eqnarray}
when both $\rho$ and $P_i$ are measured in the orthonormal frame, indicated by the hats on the indices.\footnote{The orthonormal frame is what is seen by an ideal observer within the spacetime, who observes that space is locally flat. Mathematically, the transformation that brings the stress-energy tensor into the orthonormal frame also transforms the metric into the Minkowski metric, $\eta = $diag(-1,1,1,1). This orthonormality aids with WEC analysis, since the maxmum velocity any timelike observer can have in any direction is simply 1.}

Normal matter obeys the WEC, so metrics which are derived by solving the Einstein equations for a realistic stress-energy tensor $T^{\hat{\mu}\hat{\nu}}$ automatically comply with Eq.~\eqref{eq:wec}.  However, the purpose of this paper is to demonstrate that metrics with the similar orbital mechanics can produce very different (yet WEC-compliant) $T^{\hat{\mu}\hat{\nu}}$. (Sec. \ref{sec:results}.) Therefore, the metrics presented here are created from orbital mechanics, not preexisting $T^{\hat{\mu}\hat{\nu}}$, and the $T^{\hat{\mu}\hat{\nu}}$ presented here are created from the metrics by assuming that the Einstein equations hold true\footnote{See \cite{carroll} Ch 4.6 for a discussion of this approach to GR problems.}.  The presented metrics and their associated $T^{\hat{\mu}\hat{\nu}}$  are all automatically solutions to the Einstein equations, because the $T^{\hat{\mu}\hat{\nu}}$ were formulated that way. The challenge is then to show that the $T^{\hat{\mu}\hat{\nu}}$ do not violate realism conditions like the WEC, and therefore that the metrics are valid spacetime configurations.

\subsection{Finding a form for $g_{tt}$ which produces constant-speed circular orbits at high $r$ }\label{sec:gtt}

Let the speed of a stable circular orbit be represented by $V_{tan}(r)$. A spacetime metric which produces constant-speed circular orbits at high $r$ has:

\begin{equation}
\lim_{r \to \approx \infty} V_{tan}(r) = constant = v_{tan}
\label{eq:limit}
\end{equation}

The lower-case $v_{tan}$ is a constant, and it indicates the limit of the r-dependent orbital speed, $V_{tan}(r)$. $r \to \approx \infty$ is used to indicate that we are interested in the limit at very high $r$, though not necssarily $\infty$ itself.

Let's modify a Schwarzchild metric to produce a constant $v_{tan}$ as $r \to \infty$. Consider spacetime metrics which have line elements of the form

\begin{equation}\label{eq:metric-raw}
ds^2 =  - \left( 1-\frac{2M}{r} + h(r) \right) dt^2 + \frac{1}{ 1-\frac{2M}{r} + f(r) } dr^2 + r^2 d\theta^2 + r^2 \sin^2{\theta}  d\phi^2
\end{equation}

Eq.~\eqref{eq:metric-raw} is basically the Schwarzchild metric with 2 extra terms, $f(r)$ and $h(r)$.  $M$ is a constant, representing the mass of a Schwarzchild black hole centered at $r=0$. Note that the unmodified Schwarzchild metric is a vacuum metric: There is no mass outside the event horizon. In this modified Schwarzchild metric, all mass-energy outside the event horizon will depend on $f(r)$ in the orthonormal frame (see Eq.~\ref{eq:rho-basic-generic}), and pressure will depend on both $f(r)$ and $h(r)$ (see Appendix \ref{app:full-Tmv}). $f$ and $h$ ought to be formulated to produce the desired $V_{tan}(r)$ and realistic energy conditions in the entire spacetime. 

The line element in Eq.~\eqref{eq:metric-raw} can be used to create a Lagrangian, from which we can extract the geodesics of the metric:
\begin{equation}
\mathcal{L} = \frac{1}{2} \left(- \left( 1-\frac{2M}{r} + h(r) \right) \dot{t}^2   + \frac{1}{ 1-\frac{2M}{r} + f(r)} \dot{r}^2 + r^2\dot{\theta}^2 + r^2\sin{\theta}^2\dot{\phi}^2 \right)
\label{eq:lagrangian}
\end{equation}

The dotted coordinates are derivatives in $s$, e.g. $\dot{t} = \frac{dt}{ds}$. The factor of 1/2 is aesthetic and will drop out of the geodesics; it is included here to cancel the factors of 2 that come from taking the derivative of squared coordinates. 

We are interested in the radial acceleration associated with gravitational attraction, i.e. the spontaneous  acceleration $\frac{\ddot{r}}{\dot{t}^2}$. Since this is a spherically symmetric metric, we can choose to consider only the equatorial plane with no loss of generality ($\theta = \frac{\pi}{2}$; $\dot{\theta} = 0$).\footnote{As noted elsewhere in the text, the dark pressure metric produces anisotropic pressures, but these pressures are spherically symmetric (like the metric itself). The `equatorial plane' is therefore identical to every other plane that passes through the origin.  }  

To find the spontaneous radial acceleration, solve $\frac{d} {ds}\frac{\partial\mathcal{L}}{\partial\dot{r}} = \frac{\partial\mathcal{L}}{\partial{r}}$.  As is usual for Lagrangians, treat $r$ and $\dot{r}$ as completely independent variables.

Start with the left side:
\begin{equation}
\frac{\partial\mathcal{L}}{\partial\dot{r}} = \frac{1}{ 1-\frac{2M}{r} + f(r)} \dot{r}
\label{eq:solvingL-lhs-1}
\end{equation}
\begin{equation}
\frac{d} {ds}\frac{\partial\mathcal{L}}{\partial\dot{r}} =  \frac{1}{ 1-\frac{2M}{r} + f(r)} \ddot{r} - \frac{ \frac{4M}{r^2} + \frac{df}{dr} }{\left( 1-\frac{2M}{r} + f(r)\right)^2}  \dot{r}^2
\label{eq:solvingL-lhs-2}
\end{equation}
If we are interested only in bodies in circular orbits, $\dot{r} = 0$, and therefore the second term of Eq.~\eqref{eq:solvingL-lhs-2} drops out.

Now do the right side. At $\theta = \frac{\pi}{2}$, $\ddot{\theta} = 0$: 
\begin{equation} \label{eq:solvingL-rhs}
\frac{\partial\mathcal{L}}{\partial{r}} = -\left(\frac{M}{r^2} + \frac{1}{2}\frac{dh}{dr}\right) \dot{t}^2 - \frac{1}{2}\frac{ \frac{4M}{r^2} + \frac{df}{dr} }{(1-\frac{2M}{r} + f(r))^2}  \dot{r}^2 + r \dot{\phi}^2
\end{equation}
Again, if we are interested only in bodies in circular orbits, $\dot{r} = 0$, and therefore the $\dot{r}^2$ term will drop. We put Eqs.~\eqref{eq:solvingL-lhs-2} \& \eqref{eq:solvingL-rhs} together to get:
\begin{equation}
 \frac{1}{ 1-\frac{2M}{r} + f(r)} \ddot{r} = -\left(\frac{M}{r^2} + \frac{1}{2}\frac{dh}{dr}\right) \dot{t}^2+ r \dot{\phi}^2
\end{equation}
Rearrange to get the radial acceleration due to gravity:
\begin{equation}\label{eq:radial-acceleration}
\frac{ \ddot{r}}{\dot{t}^2} = \left(1-\frac{2M}{r} + f(r)\right)\left(-\left(\frac{M}{r^2} + \frac{1}{2}\frac{dh}{dr}\right)+ r \frac{\dot{\phi}^2}{ \dot{t}^2}\right)
\end{equation}

Eq.~\eqref{eq:radial-acceleration} has a spontaneous acceleration term, $-\left(\frac{M}{r^2} + \frac{1}{2}\frac{dh}{dr}\right)$. The $-\frac{M}{r^2}$ is a Newtonian gravitation term. There is also a term that depends on $\frac{\dot{\phi}}{ \dot{t}}$, i.e. the angular speed in rad/s. This term reflects centrifugal acceleration that occurs when angular speed is not zero. The centrifugal acceleration term can be written in terms of $V_{tan}(r)$: 

\begin{equation}
r \frac{\dot{\phi}^2}{ \dot{t}^2} = \frac{V_{tan}^2(r)}{r}
\end{equation}
 
Note that to continue to move in a circular orbit, a body must have a total radial acceleration of 0. This means that
\begin{equation}
\frac{V_{tan}^2(r)}{r} = \frac{M}{r^2} + \frac{1}{2}\frac{dh}{dr}
\end{equation}
so
\begin{equation} \label{eq:this-must-be-true}
V_{tan}^2(r) = \frac{M}{r} + \frac{r}{2}\frac{dh}{dr}
\end{equation}

We want $V_{tan}^2(r) \rightarrow v_{tan}^2$ at large $r$, to comply with Eq.~\eqref{eq:limit}. Naively, one way to make this true of Eq.~\eqref{eq:this-must-be-true} is to set $\frac{dh}{dr} = 0$ (so $h$ is a constant and possibly zero), and assume $M$ is directly proportional to $r$, i.e. $M = M_0 r$, with $M_0$ constant.  If that were a viable operation, it would imply that galactic mass grows linearly with distance (and therefore that there is a large unseen $\frac{1}{r^2}$ density of mass in space).  However, that is mathematically shady, because we derived the geodesics treating $M$ as a constant. 

If we continue to respect our earlier treatment of $M$ as constant, then $\frac{M}{r}$ must go to 0 as $r$ goes to $\infty$.  Therefore, the form of $V_{tan}^2(r)$ at large $r$ will depend on $h(r)$, via the term $\frac{r}{2}\frac{dh}{dr}$.

Since $h(r)$ depends only on $r$, this can be set up as a simple ordinary differential equation, assuming that $r$ has gotten so large that $\frac{M}{r} \approx 0$ and $V_{tan}^2(r) \approx v_{tan}^2 $:
\begin{eqnarray}
v_{tan}^2 = \frac{r}{2}\frac{dh}{dr} \nonumber \\
\Rightarrow 2 v_{tan}^2 \frac{1}{r}dr = dh \nonumber \\
\Rightarrow h(r) = 2 v_{tan}^2 \log{r} + \texttt{a constant of integration}
\end{eqnarray}

The constant of integration can be chosen so that $h = 0$ at small $r$, preventing $h(r)$ from exerting noticeable affects at small $r$. Namely, choose the constant of integration to be  $-2 v_{tan}^2 \log{2M}$, so that
\begin{equation}\label{eq:h-solution}
 h(r) = 2 v_{tan}^2 \log{\frac{r}{2M}}
\end{equation}

The inclusion of the $2M$ ensures $h=0$ at $r=2M$, the event horizon of the black hole in the galactic center, so that $h(r)$ will not interfere with the black hole itself.

The final result for the  tangential velocity of a circular orbit is:

\begin{equation}
V_{tan}(r) = \sqrt{\frac{M}{r} + v_{tan}^2}
\label{eq:vtan}
\end{equation}

\begin{figure}
\includegraphics[width=\textwidth]{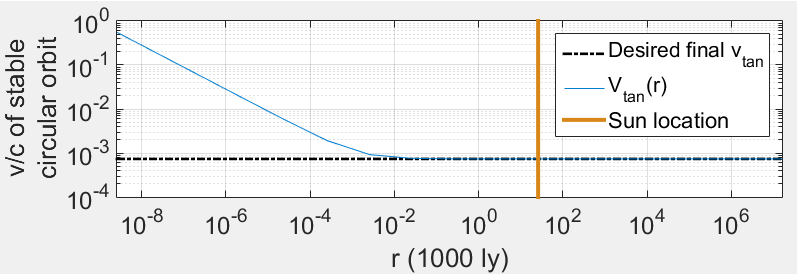}
\caption{Log-Log plot of $V_{tan}(r)$ (Eq.~\eqref{eq:vtan}), as derived from the metric in Eq.~\eqref{eq:metric-raw} and requirement of Eq.~\eqref{eq:limit}. Here, $M = 25$ light-seconds (equivalent to about 5 million suns, the approximate mass of Sagitarius A*), and $v_{tan} = 7.34\times 10^{-4}$ (about 220 km/s).  $V_{tan}(r)$ approaches the desired constant value $v_{tan}$ at high $r$.}
\label{fig:V_of_r}
\end{figure}

$V_{tan}(r)$ is plotted in Fig.~\ref{fig:V_of_r}, using  $v_{tan} = 7.34\times 10^{-4}$ (about 220 km/s) and $M=25$ (units of light-seconds, such that this $M$ corresponds to the mass of a Schwarzchild black hole with radius $2M = 50$ light-seconds.)  Note that this $V_{tan}(r)$ does not reproduce galactic rotation curves at low $r$, because it is a simply-modified Schwarzchild metric that contains only a central black hole and a nonspinning, spherically-distributed medium. A metric that can reproduce galactic rotation curves at low $r$ will have a more complicated form (see \cite{cooperstock}).

The $V_{tan}(r)$ of Eq.~\ref{eq:vtan} is similar to a Keplerian velocity plus a constant term. This is by design: As discussed in Sec.~\ref{sec:math-basis}, rotation curves like the one plotted in Fig.~\ref{fig:V_of_r} are a starting point in our analysis, which will produce a stress-energy tensor as an output.  This `reverse engineering' approach allows us to consider, in isolation, whether mass is the only significant contributor to rotation curves, or whether other components of the stress-energy tensor can be important to orbital dynamics.

Using the $h(r)$ of Eq.~\eqref{eq:h-solution}, the form of $g_{tt}$ is:

\begin{equation}
g_{tt}  = - \left( 1-\frac{2M}{r} + 2v_{tan}^2 \log{\frac{r}{2M}}\right) 
\label{eq:gtt}
\end{equation}

\subsection{Proposing a form for $g_{rr}$.}\label{sec:grr}

The component $g_{tt}$ alone is not energetically meaningful. We also need a form for $g_{rr}$.  $g_{rr}$ controls the mass-energy density $\rho$ in a spherically symmetric, static spacetime. For the generic, r-dependent metric of Eq.~\ref{eq:generic-metric}, the mass-energy density $\rho$ in orthonormal coordinates is given by:

\begin{equation}
\rho = \frac{1}{8\pi} \left( \frac{g_{rr}^2 - g_{rr} + r\frac{d}{dr}g_{rr}}{r^2 g_{rr}^2} \right)
\label{eq:generic-rho}
\end{equation}  

This $\rho$ is found by first computing the Einstein tensor $G_{\mu\nu}$ associated with the metric, then assuming that $G_{\mu\nu}$ is an accurate representation of the curvature induced by some stress energy tensor; i.e. the Einstein equation $G_{\mu\nu} = 8\pi T_{\mu\nu}$ is assumed to be true. $T_{\mu\nu}$ is then orthonormalized with a transformation matrix $\Lambda$, such that $\Lambda^T g_{\mu\nu} \Lambda$ equals the cartesian Minkowski metric, and $\Lambda^T T_{\mu\nu} \Lambda = T_{\hat{\mu}\hat{\nu}}$.  The raised-index $T^{\hat{\mu}\hat{\nu}}$ has a simple relation to the lowered-index $T_{\hat{\mu}\hat{\nu}}$:  For spatial indices $i,j$ and time index 0, $T_{\hat{0}\hat{0}} = T^{\hat{0}\hat{0}} = \rho$, and $T_{\hat{i}\hat{j}} = T^{\hat{i}\hat{j}}$, but $T_{\hat{0}\hat{i}} = -T^{\hat{0}\hat{i}}$.

For the metric we are constructing, let $g_{rr}$ depend on the function $f(r)$ as shown in Eq.~\ref{eq:metric-raw}. Then $\rho$ has a simple relation to $f(r)$:

\begin{equation}
\rho = \frac{1}{8\pi}\left( -\frac{r\, \frac{\partial}{\partial r} f\!\left(r\right) + f\!\left(r\right)}{r^2}\right)
\label{eq:rho-basic-generic}
\end{equation}

$\rho$ is unknown at this point, but it should decrease with rising $r$, since we expect the galaxy to be less dense at high $r$. Consider a form for $f(r)$:

\begin{equation}\label{eq: f-solution}
f(r) = -\frac{h(r)}{B}= -\frac{2 v_{tan}^2 }{B}\log{\frac{r}{2M}}  
\end{equation}

where $B$ is an arbitrary constant. This $f(r)$ leads to an acceptable\footnote{For now, assume it is acceptable for $|g_{tt}|$ and $g_{rr}$ to not perfectly equal 1 at cosmological scales; this will be addressed in Sec. \ref{sec:FRW-mod}.} result for the stress energy tensor when input with realistic values for $M$, $v_{tan}$, and $r$, and various $B$. It will be seen that $B$ is a scaling constant that is inversely proportional to $\rho$ (Eq.~\ref{eq:rho}).\footnote{This $f(r)$ has no derivation. It was discovered via experimentation.}\,\footnote{The choice of the letter `B' for the density scaling constant is inspired by `baros', Greek for `weight'.}

 The form of the metric with this $f(r)$ and the $g_{tt}$ found in Sec.~\ref{sec:gtt} is:
\begin{equation}\label{eq:metric-basic}
ds^2 =  -\left( 1-\frac{2M}{r} + 2 v_{tan}^2 \log{\frac{r}{2M}} \right) dt^2 + \frac{1}{ 1-\frac{2M}{r} - \frac{2}{B} v_{tan}^2 \log{\frac{r}{2M}}} dr^2 + r^2 d\Omega^2
\end{equation} 

The $T^{\hat{\mu}\hat{\nu}}$ produced by this metric is diagonal, i.e. it contains only mass-energy density $\rho$ and 3 directional pressures $P$.  $P_{\theta}$  and $P_{\phi}$ are identical in orthonormal coordinates. We use $P_{\Omega}$ to represent them both. 

$\rho$ has a simple form. For $L = \mathrm{log}\!\left(\frac{r}{2\, M}\right)$:
\begin{equation}
\rho = \left(\frac{L+ 1}{4 \pi B\, r^2}\right)\, v^2 
\label{eq:rho}
\end{equation} 

Whereas $P_r$ is more complicated: 

\begin{eqnarray}
\label{eq:Pr} 
P_r = & \\\nonumber 
& - \frac{\left(4\, M\, v^2 - 2\, r\, v^2 + 4\, L\, M\, v^2\right)\,B + 4\, L^2\, r\, v^4 + 2\, L\, r\, v^2 + 4\, L\, r\, v^4}{8 \pi B\, \left(2\, L\, r^3\, v^2 + r^3 - 2\, M\, r^2\right)} 
\end{eqnarray}

$P_{\Omega}$ is not the same as $P_r$, and its analytic expression is too long to reproduce here. (It is given in Appendix~\ref{app:full-Tmv}.) These pressures are spherically symmetric, but anisotropic, nontrivial, and pervasive throughout the spacetime. We have called this a `dark pressure' metric because of these pressures and their ability to serve a similar function to dark matter.

Note that the mass-energy density $\rho$ in Eq.~\ref{eq:rho} is inversely scaled by $B$, a factor that does not appear anywhere in the fomula for $V_{tan}(r)$. This means that increasing $|B|$ reduces the mass-energy density of the dark pressure spacetime, without affecting the speed of stable circular orbits.  (However, $B$ is constrained by energetic realism considerations, as discussed in Sec.~\ref{sec:results}.) 

The stress-energy tensor $T^{\hat{\mu}\hat{\nu}}$ as a whole should still meet the WEC in order to be realistic. WEC evaluation is performed in Sec.~\ref{sec:results}, using both analytic and numeric techniques.  It is found that $T^{\hat{\mu}\hat{\nu}}$ meets the WEC for a broad range of $B$.


The next section will modify the dark pressure metric to produce isotropic pressure at high $r$, in keeping with cosmological observations.

\subsection{Merging the dark pressure metric with a cosmological metric} \label{sec:FRW-mod}

 The unmodified `Basic' dark pressure metric is not asymptotically isotropic or asymptotically flat, although its discrepancy from flatness may be very small at cosmological $r$, depending on the value of $v$.  If we want the metric to give realistic features at high $r$, it must be modified. Rather than `flattening' the metric at cosmological $r$, we choose to merge it with a cosmological metric instead. We choose a modification based on the Robertson-Walker (RW) metric. The RW metric has the form:
  
  \begin{equation}
  	ds^2 = -dt^2 + a^2(t)\left( \frac{1}{1-\kappa r^2}dr^2 + r^2 d\Omega^2\right)
	\label{eq:RW}  
  \end{equation}

Here, $a(t)$ is the scale factor of the expanding universe at a given time $t$, and $\kappa$ is the un-normalized curvature of the universe. This RW metric is also a `Friedmann-Robertson-Walker' (FRW) metric if it produces a stress-energy tensor which obeys the Friedmann equations (\cite{carroll}, Eq.~8.67, 8.68):

  \begin{equation}
  	\left( \frac{\dot{a}}{a}\right)^2 =  \frac{8\pi G}{3c^4}\rho - \frac{\kappa}{a^2} 
	\label{eq:friedmann-1}  
  \end{equation}
  
  \begin{equation}
  	\frac{\ddot{a}}{a} = -\frac{4\pi G}{3c^4}\left( \rho + 3P \right)
	\label{eq:friedmann-2}  
  \end{equation}  
  
The factors of $c$ in both equations are necessary if inputting numeric $G$,$\rho$, and $P$ in SI units.  See Appendix~\ref{app:units}..
  
The basic dark pressure metric of Eq.~\ref{eq:metric-basic} can be modified to mimic the form of an RW metric. To do this, multiply all spatial components of the metric by the scale factor $a(t)$, and include a term of $(-\kappa r^2)$ in the denominator of the $g_{rr}$ component. This procedure yields the complete modified dark pressure metric: 

\begin{equation}\label{eq:metric-FRW-mod}
ds^2 =  -\left( 1-\frac{2M}{r} + 2 v_{tan}^2 \log{\frac{r}{2M}} \right) dt^2 + a^2(t) \left( \frac{1}{ 1-\frac{2M}{r} - \frac{2}{B} v_{tan}^2 \log{\frac{r}{2M}} - \kappa r^2} dr^2 + r^2 d\Omega^2 \right)
\end{equation}

It is important to note that this metric has no junctions: We have not pasted the Basic metric onto an RW metric at some location in spacetime. Instead, the metric of Eq.~\ref{eq:metric-FRW-mod} continuously covers the entire spacetime, with the property that it resembles the Basic metric at low $r$ and an RW metric at high $r$.  We will refer to this as the `FRW Mod' metric, since it (nearly) obeys the standard Friedmann equations for realistic input parameters. (The `nearly' is analytically expected and reasonable, see following discussion.)

The FRW Mod metric has more complicated geodesics than the Basic metric, but the speed of stable circular orbits has almost the same form as Eq.~\ref{eq:vtan}. Assuming that the orbiting bodies are not significantly drifting away from the galactic center due to expansion, we again want $\dot{r}=0$, $\ddot{r}=0$. We obtain:

\begin{equation}
V_{tan}(r) = \frac{1}{a(t)}\sqrt{\frac{M}{r} + v_{tan}^2} \texttt{   for complete metric}
\label{eq:vtan-complete}
\end{equation}

This implies that the RW modification to the metric does not significantly affect $V_{tan}(r)$ of the central galaxy, as long as the scale factor $a(t)$ is nearly constant over the timescale of our observations\footnote{At a given time, we can arbitrarily set the scale factor to 1, so if the scale factor is effectively constant over the timescale of our observations, the RW modification has effectively no impact on $V_{tan}(r)$.}. 

The FRW Mod metric does not precisely meet the Friedmann equations at cosmological $r$, because the Friedmann equations are derived by expressing the Ricci tensor of the metric of Eq.~\ref{eq:RW} as a solution to the Einstein equations (see \cite{carroll}, Sec.~8.2, 8.3):

\begin{equation}
\label{eq:fried-root}
R_{\mu\nu} = 8\pi ( T_{\mu\nu} -\frac{1}{2} g_{\mu\nu}T )
\end{equation}

where $T$ is the contraction of $T_{\mu\nu}$.  Given that we construct $T_{\mu\nu}$ by assuming the Einstein equations hold true, this formula is automatically true for the FRW Mod metric. However, it does not produce Friedmann-style equations exactly like Eqs.~\ref{eq:friedmann-1}~\&~\ref{eq:friedmann-2}. This is expected, since the FRW Mod metric is not identical to the RW metric. The question is whether the difference between Eqs.~\ref{eq:friedmann-1}~\&~\ref{eq:friedmann-2}  and the FRW Mod's Friedmann-style equations  is numerically distinguishable: In other words, if a dark pressure galaxy existed in our universe, would it have any noticeable effect on cosmological observations?  If not, that is a point in favor of the dark pressure galaxy as a viable spacetime configuration, because it would not disrupt existing cosmological models.

To get an analytic sense of how closely the FRW Mod metric complies with the standard Friedmann equations, we can naively plug the FRW Mod's $\rho$ into the righthand side of Eq.~\ref{eq:friedmann-1}.  If we set $a(t_0) = 1$ and use $G=c=1$, we get

\begin{equation}
\frac{8\pi}{3}\rho - \kappa = \frac{\dot{a}^2\,r}{2\,L\,r\,v^2-R+r}-\kappa+\frac{2\,L\,v^2+2\,v^2+3\,B\,\kappa\,r^2}{3\,B\,r^2}
\label{eq:fried1-rhs}
\end{equation}

The limit of Eq.~\ref{eq:fried1-rhs} as $r$ becomes very large is $\frac{\dot{a}^2}{2\,L\,v^2+1}$ (assuming $L$ changes much more slowly than $r$). Numerically, this is essentially indistiguishable from the lefthand side of Eq.~\ref{eq:friedmann-1}, because for $v << c$ and cosmological $r$:

\begin{equation}
\dot{a}^2 \approx \frac{\dot{a}^2}{2\,L\,v^2+1}
\end{equation}

Numeric checks on the FRW Mod's compliance with the standard Friedmann equations will be described in Sec.~\ref{sec:results} and Table~\ref{tab:fried}. Results indicate that the existence of a dark pressure galaxy would not conflict with FRW cosmology for reasonable input parameters.

\section{Results}\label{sec:results}
For both the Basic and FRW Mod versions of the metric, the resulting orthonormal stress-energy tensor $T^{\hat{\mu}\hat{\nu}}$ should meet the WEC in order to be realistic. Additionally, the FRW Mod metric should (nearly) obey the standard Friedmann equations at high $r$. The Basic metric should\footnote{The GTOV equation was derived in \cite{Riazi_2016} using assumptions that do not apply to the dark pressure metric, but the resulting GTOV equation is nonetheless valid for the Basic dark pressure metric, see Sec.~\ref{sec:gtov}.} obey the Generalized Tolman-Oppenheimer-Volkoff (GTOV) equation for hydrostatic equilibrium, if it is to be considered a static solution to the Einstein Field Equations for a fluid.\footnote{The FRW Mod metric features cosmological expansion, so cannot be said to be in equilibrium. However, note that the FRW Mod metric produces an almost identical stress-energy tensor to the Basic metric at galactic-scale $r$. Therefore, if the Basic metric obeys the GTOV, the FRW Mod metric almost obeys the GTOV at galactic-scale $r$ and sufficiently small timescales, i.e. it has local equilibrium.}
\subsection{Analytic WEC Compliance of the Basic metric}\label{sec:analytic-basic}

The formulas for $T^{\hat{\mu}\hat{\nu}}$ can be produced analytically from the metric, using a symbolic software package (see Appendix~\ref{app:full-Tmv} for the formulas for all nonzero components of the Basic $T^{\hat{\mu}\hat{\nu}}$).   The Basic $T^{\hat{\mu}\hat{\nu}}$ can partially be checked for WEC compliance analytically.  Using $R = 2M$ and $L=\log\left(\frac{r}{R}\right)$:

\begin{itemize}
\item The formula for mass-energy density $\rho$ (Eq.~\ref{eq:rho}) is always positive for $B>0$ and $r \geq 2M$ (i.e. outside the event horizon of the central black hole), therefore the $\rho \geq 0$ requirement of the WEC is met.

\item The formula for $\rho + P_r$ is $\frac{v^2\,\left(B+1\right)\,\left(r-R-L\,R\right)}{4\,B\,r^2\,\pi \,\left(2\,L\,r\,v^2-R+r\right)}$.  Of these factors, all are positive for $B>0$, except possibly $(r-R-RL)$. Call this factor $W$. At $r = R$, $W= 0$, so the question is whether $W$ ever becomes negative for $r > R$.  Examining the derivative $\frac{dW}{dr} = 1 - \frac{R}{r}$, we see that $\frac{dW}{dr}$ is always positive for $r>R$, and therefore $W$ increases monotonically outside the event horizon. This means that the $\rho + P_r \geq 0$ requirement of the WEC is met.

\item  The WEC also requires that $\rho + P_{\Omega} \geq 0$, and this cannot be evaluated purely analytically. The full expression for $\rho + P_{\Omega}$ is given in Appendix~\ref{app:full-Tmv}.  Fundamentally, $\rho + P_{\Omega} = A(BX + Y)$, where $A$ is always positive for positive $B$, and $X$ and $Y$ are long sums, both with maximum order $r^2$.  Whether or not $\rho + P_{\Omega} \geq 0$ depends on $B$, and therefore the WEC can be seen as constraining $B$.   It is noted in the Appendix that the highest-order term in $X$ is negative.  This means that $X$ is negative for sufficiently high $B$ and $r$, i.e. making $B$ too high may cause a WEC violation at high $r$. The question is then whether this WEC violation occurs at a valid $r$:  A very high $B$ that causes a WEC violation in local space is clearly a problem, whereas a lower $B$ that causes a WEC violation outside the cosmological event horizon is less concerning, because the Basic metric isn't applicable at such distances. This is part of the motivation for presenting the FRW Mod version of the metric and a corresponding numeric analysis: The FRW Mod metric is relevant up to cosmological scales and achieves constant, isotropic, WEC-compliant pressures, whereas the Basic metric is badly behaved if extended far beyond a reasonable range of $r$.\footnote{Very high $B$ values also lead to positive pressures at low $r$ which are well in excess of $\rho$.}  Note, however, that the FRW Mod would be unable to prevent WEC violations at sub-cosmological $r$ if $B$ were set excessively high. 

At the lower limit $r\rightarrow (R=2M)$, we have $\rho + P_{\Omega} = \frac{v^2\,\left(8\,v^2+3\,B+7\right)}{32\,B\,R^2\,\pi \,\left(2\,v^2+1\right)}$, a positive number for $B>0$, so we can say that there is no WEC violation at the event horizon itself.


\end{itemize}

The $\rho$, $P_r$, and $P_{\Omega}$ of the Basic metric are also evaluated numerically in the next section. This is done to compare them with the equivalent values for the FRW Mod metric. Additionally, it illustrates that the Basic metric produces no WEC violations within the cosmic event horizon over a wide range of $B$.

\subsection{Input parameters for numeric analysis.}\label{sec:inputs}

The FRW Mod metric's stress-energy tensor is too complicated for pure analytic assessment, since it contains messy factors of $a(t)$, $\dot{a}(t)$, $\ddot{a}(t)$, and $\kappa$. Therefore, the energetic realism of the FRW Mod metric is best evaluated numerically. 

Results will be given in SI units and kilo-lightyears (kly), but computation is done in geometerized units, where all values are expressed in meters.

For all analysis of $T^{\hat{\mu}\hat{\nu}}$, we use two values based on the Milky Way, three arbitrary positive values for $B$, and an exponential range of values for $r$:

\begin{itemize}\label{item:vals}
\item Use  $M = 25c$ ($\approx 7.49 \times 10^9$ meters in geometerized units), half the radius of a black hole with a mass of $10^{37}$ kg, about 5 million suns.  This is comparable to the mass of the black hole in the center of the Milky Way, Sagitarius A*.

\item Use $v_{tan} = 220$ km/s, an estimate of galactic orbital speed at high $r$.  In geometerized units, this is $v_{tan} = 7.34\times10^{-4}$.

\item Use three different $B$ values to see how $B$ affects $T^{\hat{\mu}\hat{\nu}}$ and WEC compliance: $B = 1,10,100$.

\item Use $r$ ranging from slightly outside the central event horizon ($r_1 \approx 102$ light-seconds $\approx 3.06 \times 10^{10}$ m $\approx 3.24\times 10^{-9}\,\texttt{ kly}$, about $4M$) to the cosmic event horizon ($r_{CEH} = 16,000\,\texttt{kly} = 1.51\times10^{26}$ meters). Include the Sun's location at $r_{Sun} = 26\,\texttt{kly} = 2.46\times10^{20}$ meters. The $r$ used in the tables increase by a factor of 26 at each step, except for the last step, which is $r_{CEH}$.
\end{itemize}

The FRW Mod has additional input parameters:  $a(t)$, $\dot{a}(t)$, $\ddot{a}(t)$, and $\kappa$,  and furthermore has a time dependence.  We limit our analysis to $t=t_0$, the present time.  $a(t_0)$ can arbitrarily be set to 1, but $a(t)$ itself is not a known function. Its derivatives $\dot{a}$ and $\ddot{a}$ are not known for all time, and cannot be set arbitrarily. However, $\dot{a}(t_0)$ and $\ddot{a}(t_0)$ can be estimated from cosmological observations, and this is an active area of research (\cite{planck},\cite{Camarena_2020}). We use realistic demonstrative values for $a(t_0)$, $\dot{a}(t_0)$, $\ddot{a}(t_0)$, and $\kappa$ to show that Eq.~\ref{eq:metric-FRW-mod} is asymptotically a cosmological metric. However, other input values will yield qualitatively similar results.  A full cosmological evaluation is outside the scope of this work.

 $\dot{a}(t_0)$ is related to the Hubble parameter $H_0$ via the equation $H = \frac{\dot{a}}{a}$  (\cite{carroll}, Eq.~8.60), where $H_0 = H(t_0)$. $\ddot{a}(t_0)$ is related to a `deceleration parameter' $q_0$ via the equation $q = -\frac{a \ddot{a}}{\dot{a}^2}$ (\cite{carroll}, Eq.~8.73), where $q_0 = q(t_0)$. This work uses values of $H_0$ and $q_0$ given in Camarena and Marra (\cite{Camarena_2020}, Eqs.~B1, B2): $H_0 = 74.62 km/s/MPc \approx 8.0665 \times 10^{-27} \,/m$, $q_0 = -0.90$, and a scale factor $a(t_0) = 1$. Together, these values yield numeric values for $a$ and its derivatives:

\begin{eqnarray}\label{eq:a-vals}
a(t_0) = 1 \nonumber \\
\dot{a}(t_0) \approx 8.0665\times 10^{-27} \, m^{-1} \nonumber \\
\ddot{a}(t_0) \approx 5.8561 \times 10^{-53}  \,  m^{-2}
\end{eqnarray}

$\kappa$ is also unknown. Cosmological observations indicate that the universe is very flat, and therefore that its curvature $\kappa$ is very close to zero. If $\kappa$ is nonzero, it may be positive or negative. It is has the formula $\kappa = \pm R_0^{-2}$ (\cite{carroll}, Eq.~8.42), where $R_0$ is the radius of curvature of the universe at $t_0$.

This work will perform numerical analysis using a value of $R_0 = 1000\,Gly = 9.46 \times 10^{27}$ meters, much larger than the observable universe. This value is somewhat arbitrary, but within the constraints on universal curvature suggested by the Planck collaboration \cite{planck}. It yields $\kappa \approx 1.117 \times 10^{-56} \,m^{-2}$. 

Of all these numeric values, only $R_0$ and $B$ were chosen arbitrarily. $R_0$ is simply a very large round number, and the 3 values of $B$ are intended to span a range of $B$ while producing numeric results for $\rho$ that are easy to compare across $B$ (since $\rho$ is inversely proportional to $B$, changing $B$ by a factor of 10 leaves the digits of $\rho$ unchanged). The exponential grid of $r$ values was designed to get close to the event horizon while also including the location of the Sun, where the density of the interstellar medium is best known and can be meaningfully compared to the numeric output $\rho$. The final $r$ value breaks the exponential pattern, but this is because the next exponential value of $r$ lands outside the cosmic event horizon.  All other numerics are standard rounded or published values; no `hand-tuning' was involved.

Numeric values for $T^{\hat{\mu}\hat{\nu}}$ must be computed in geometerized units, where $G=c=1$ and all units are expressed relative to meters. However, results will be given in SI units and kilo-lightyears ($kly$). See Appendix \ref{app:units} for unit conversions.

\subsection{Numeric values for $T^{\hat{\mu}\hat{\nu}}$: WEC and Friedmann compliance }\label{sec:numeric}

Numeric values for $T^{\hat{\mu}\hat{\nu}}$ at $B=1, 10, 100$ are shown in Tables \ref{tab:Tmv-B1}, \ref{tab:Tmv-B10},\ref{tab:Tmv-B100}\footnote{Due to the wide range of scales and mixed signs of the  $T^{\hat{\mu}\hat{\nu}}$ values, this data cannot be easily plotted.}. These numbers are outputs created by inputting the parameters of Sec.~\ref{sec:inputs} to the exact analytic equations of $T^{\hat{\mu}\hat{\nu}}$ (which were obtained by requiring that $T^{\hat{\mu}\hat{\nu}}$ solve the Einstein equations for the metrics of Eqs.~\ref{eq:metric-basic} and \ref{eq:metric-FRW-mod}; all equations for the Basic metric are given in Appendix~\ref{app:full-Tmv}). The numeric results for the Basic metric are shown in the `Basic' columns; the results for the FRW Mod metric are shown in the `FRW Mod' columns.  

For the Basic metric, the $T^{\hat{\mu}\hat{\nu}}$ appear to meet WEC everywhere for all evaluated $B$ and $r$ (see Eq.~\ref{eq:wec} for WEC formula). From Sec.~\ref{sec:analytic-basic}, we know that only the  $P_{\Omega}$ component of the Basic $T^{\hat{\mu}\hat{\nu}}$ is analytically capable of producing a WEC violation, and that it does not do so at the inner limit $r=2M$.  We also know that both $P_{\Omega}$ and  $\rho + P_{\Omega}$ are roughly proportional to second-order polynomials of $r$ (i.e., their signs cannot fluctuate many times), and that any WEC violation would depend on the numeric values of $B$ and $r$. In Tables \ref{tab:Tmv-B1}, \ref{tab:Tmv-B10}, \ref{tab:Tmv-B100}, we see that  $P_{\Omega}$ is over an order of magnitude smaller than $\rho$ for all evaluated $B$ and $r$ in the Basic metric.  This means it is unlikely there is a `missed' WEC violation for any $r$ between $r=2M$ and the cosmic event horizon, since any such violation would require a large fluctuation of the relative values of $P_{\Omega}$ and $\rho$ between one evaluated $r$ and the next.

In all Basic cases, $\rho$ is inversely proportional to $B$, as expected from Eq.~\ref{eq:rho}.  The digits of $\rho$ are the same in every Table because the $B$ values differ by a factor of 10, so only the exponents differ. All these metrics share the same speed for stable circular orbits (Sec. \ref{sec:gtt}). This indicates that it is energetically realistic under the WEC for a galaxy to have rotation curves that are not due solely to the mass of the galaxy.

By contrast, the pressure values do not have a clean dependence on $B$.  This is expected based on their formulas (Appendix~\ref{app:full-Tmv}).

The numeric value of the FRW Mod stress-energy tensor are nearly identical to the Basic values until some $r$ on the order of $100\,kly$, where the `FRW' aspects of the FRW Mod metric begin to dominate, and $\rho$ and $P$ begin to stabilize to constant isotropic values.  There is no junction where the Basic metric is replaced by FRW;  this is smooth numerical effect. The Einstein equations remain satisfied as long as we accept that these $\rho$ and $P$ constitute a physically valid stress-energy tensor\footnote{Carroll \cite{carroll} states in Ch. 4.6 that `any metric at all' obeys Einstein's equations if you demand that $G_{\mu\nu} = 8\pi T_{\mu\nu}$; the real concern is whether the $T_{\mu\nu}$ in question represents a physically plausible configuration of stress-energy.} (which is the reason we check for WEC violations). It has been pointed out that the distance $r \approx 100\,kly$ is similar to the distance where galactic warping and flaring occurs for real galaxies; this is coincidental.  Warping and flaring are not relevant to the dark pressure toy galaxy, which is a spherical mass distribution (not a disc) that exists without interaction with any other massive bodies. The `FRW' aspects are apparent at this scale only because the Basic aspects have become so weak. As stated earlier, the FRW Mod is implemented largely to demonstrate that it is feasible for a dark pressure galaxy to exist in our universe (and therefore the notion that orbits do not depend on mass alone should be given due weight). Detailed intergalactic dynamics are out of scope for this work.
 
To put the weakness of the Basic metric at high $r$ in context: The critical density of the universe is $\rho_{crit} = \frac{3H_0^2 c^2}{8\pi G} \approx 9.40 \times 10^{-10}\,J/m^3$ for a Hubble parameter $H_0$ of 74.62 km/s/MPc. At high $r$, the $\rho$ produced by the `Basic' dark pressure metric is several orders of magnitude lower than $\rho_{crit}$.Therefore, cosmological metrics that produce $\rho \approx \rho_{crit}$ dominate the dark pressure metric at cosmological distances. This is the case for the FRW Mod metric, which produces  $\rho \approx \rho_{crit}$ at cosmological $r$.\footnote{$\rho$ at cosmological $r$ is about 0.01\% higher than $\rho_{crit}$, in keeping with our initial choice to make $\kappa$ very small and positive.}

As discussed in Sec.~\ref{sec:FRW-mod}, the FRW Mod metric is not expected to produce an exact solution to the Friemann equations, because the Friedmann equations are derived assuming that the classic Robertson-Walker metric (Eq.~\ref{eq:RW}) is an exact solution to the Einstein equations in the spacetime. However, in order for a dark pressure galaxy to be a plausible component of our universe, it ought to hew very closely to the Friedmann equations at cosmological $r$, so that its existence is not in conflict with accepted cosmological models. This was done analytically for the first Friedmann equation in Sec.~\ref{sec:FRW-mod}; we perform a numeric evaluation here.
 
We evaluated Eqs.~\ref{eq:friedmann-1} \& \ref{eq:friedmann-2} using the last rows of Tables \ref{tab:Tmv-B1}, \ref{tab:Tmv-B10},  \ref{tab:Tmv-B100} as the inputs for $\rho$ and $P$.\footnote{The $3P$ of Eq.~\ref{eq:friedmann-2} is more properly the sum of the pressures (\cite{carroll}, Ch.~8.3). The sum of the pressures was used instead of `$3P$' here.} In this way we can recalculate the Hubble parameter $H_0 = \dot{a}/a$ and the deceleration parameter $q_0 = -a\ddot{a}/\dot{a}^2$ as an output. We found that the resulting `output'  $H_0$ and $q_0$ are the same as the `input' $H_0$ and $q_0$ to an extremely tight margin of error, well within the measurement uncertainty of these values. The results are in Table~\ref{tab:fried}.\footnote{ The outputs $H_0$ and $q_0$ would be exactly the same as the input $H_0$ and $q_0$  if we rederived the Friedmann equations on the assumption that the FRW Mod metric is an exact cosmological model, but that would miss the point.} It is therefore reasonable to say that dark pressure does not require reworking cosmological models.

The $T^{\hat{\mu}\hat{\nu}}$ of the FRW-modified dark pressure metric is not purely diagonal. It contains no shear or angular momentum, but it does have a nonzero radial momentum density $mo_r$. Analytic examination reveals that $mo_r$ does have a very small dependence on $B$. However, $mo_r$ is identical for all $B$ to within about 0.001\% in the numeric analysis. With radial momentum included, the WEC requires that $\rho + P_r - 2|mo_r|\, \geq 0$ for every point.  $|mo_r| \times c$ is shown in Table \ref{tab:Tmv-mo-r}. $mo_r \times c$ is very small compared to $\rho$ for every evaluated $B$ and $r$, over 4 orders of magnitude less at its largest.  The dark pressure metric continues to meet the WEC for every evaluated $B$ and $r$ when curvature $\kappa$ is positive.

\begin{figure}[!ht]
\includegraphics[width=\textwidth]{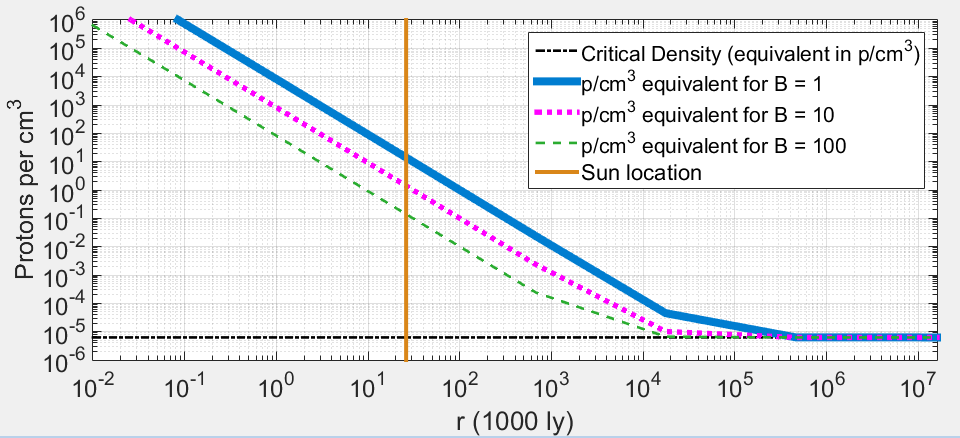}
\caption{Log-Log plot of $\rho$ for $B=1,10,100$ and the FRW Mod metric. $\rho$ has been translated into an equivalent-mass density of protons/cm\textsuperscript{3}. In the vicinity of the Sun, $B=1$ gives $\approx 14$ protons/cm\textsuperscript{3}, $B=10$ gives $\approx 1.4$ protons/cm\textsuperscript{3}, and $B=100$ gives $\approx 0.14$ protons/cm\textsuperscript{3}.}
\label{fig:protons-per-cm3}
\end{figure}

\begin{table}[hp] 
\small
\caption{Numeric values for $T^{\hat{\mu}\hat{\nu}}$ when $B=1$:}
\begin{tabular}{l|r|r|r|r|r|r}
&$Basic$&$Basic$&$Basic$&$FRW\,Mod$&$FRW\,Mod$&$FRW\,Mod$\\$r\,(kly)$&$\rho\,(J/m^3)$&$P_{r}\,(N/m^2)$&$P_{\Omega}\,(N/m^2)$&$\rho\,(J/m^3)$&$P_{r}\,(N/m^2)$&$P_{\Omega}\,(N/m^2)$\\\hline \\3.24e-09&$9.48e+15$&$-6.00e+15$&$1.88e+14$&$9.48e+15$&$-6.00e+15$&$1.88e+14$\\8.42e-08&$4.07e+13$&$-2.56e+13$&$-3.54e+12$&$4.07e+13$&$-2.56e+13$&$-3.54e+12$\\2.19e-06&$9.96e+10$&$-7.55e+10$&$-5.99e+09$&$9.96e+10$&$-7.55e+10$&$-5.99e+09$\\5.69e-05&$2.06e+08$&$-1.70e+08$&$-8.95e+06$&$2.06e+08$&$-1.70e+08$&$-8.95e+06$\\1.48e-03&$3.90e+05$&$-3.38e+05$&$-1.32e+04$&$3.90e+05$&$-3.38e+05$&$-1.32e+04$\\3.85e-02&$7.05e+02$&$-6.27e+02$&$-1.96e+01$&$7.05e+02$&$-6.27e+02$&$-1.96e+01$\\1.00&$1.23e+00$&$-1.12e+00$&$-2.90e-02$&$1.23e+00$&$-1.12e+00$&$-2.90e-02$\\26.00&$2.10e-03$&$-1.93e-03$&$-4.29e-05$&$2.10e-03$&$-1.93e-03$&$-4.29e-05$\\676.00&$3.52e-06$&$-3.27e-06$&$-6.34e-08$&$3.52e-06$&$-3.27e-06$&$-6.43e-08$\\1.76e+04&$5.82e-09$&$-5.45e-09$&$-9.38e-11$&$6.76e-09$&$-6.32e-09$&$-9.71e-10$\\4.57e+05&$9.52e-12$&$-8.96e-12$&$-1.39e-13$&$9.50e-10$&$-8.86e-10$&$-8.77e-10$\\1.19e+07&$1.54e-14$&$-1.46e-14$&$-2.05e-16$&$9.40e-10$&$-8.77e-10$&$-8.77e-10$\\1.60e+07&$8.57e-15$&$-8.11e-15$&$-1.13e-16$&$9.40e-10$&$-8.77e-10$&$-8.77e-10$\\
\end{tabular}
\label{tab:Tmv-B1}
\vspace{1em}

\caption{Numeric values for $T^{\hat{\mu}\hat{\nu}}$ when $B=10$:}
\begin{tabular}{l|r|r|r|r|r|r}
&$Basic$&$Basic$&$Basic$&$FRW\,Mod$&$FRW\,Mod$&$FRW\,Mod$\\$r\,(kly)$&$\rho\,(J/m^3)$&$P_{r}\,(N/m^2)$&$P_{\Omega}\,(N/m^2)$&$\rho\,(J/m^3)$&$P_{r}\,(N/m^2)$&$P_{\Omega}\,(N/m^2)$\\\hline \\3.24e-09&$9.48e+14$&$9.68e+14$&$1.35e+15$&$9.48e+14$&$9.68e+14$&$1.35e+15$\\8.42e-08&$4.07e+12$&$4.24e+12$&$-1.06e+11$&$4.07e+12$&$4.24e+12$&$-1.06e+11$\\2.19e-06&$9.96e+09$&$3.28e+09$&$-5.73e+08$&$9.96e+09$&$3.28e+09$&$-5.73e+08$\\5.69e-05&$2.06e+07$&$-8.81e+05$&$-8.92e+05$&$2.06e+07$&$-8.81e+05$&$-8.92e+05$\\1.48e-03&$3.90e+04$&$-9.92e+03$&$-1.32e+03$&$3.90e+04$&$-9.92e+03$&$-1.32e+03$\\3.85e-02&$7.05e+01$&$-2.74e+01$&$-1.96e+00$&$7.05e+01$&$-2.74e+01$&$-1.96e+00$\\1.00&$1.23e-01$&$-5.95e-02$&$-2.90e-03$&$1.23e-01$&$-5.95e-02$&$-2.90e-03$\\26.00&$2.10e-04$&$-1.16e-04$&$-4.29e-06$&$2.10e-04$&$-1.16e-04$&$-4.29e-06$\\676.00&$3.52e-07$&$-2.13e-07$&$-6.34e-09$&$3.53e-07$&$-2.14e-07$&$-7.22e-09$\\1.76e+04&$5.82e-10$&$-3.76e-10$&$-9.38e-12$&$1.52e-09$&$-1.25e-09$&$-8.87e-10$\\4.57e+05&$9.52e-13$&$-6.46e-13$&$-1.39e-14$&$9.41e-10$&$-8.78e-10$&$-8.77e-10$\\1.19e+07&$1.54e-15$&$-1.09e-15$&$-2.05e-17$&$9.40e-10$&$-8.77e-10$&$-8.77e-10$\\1.60e+07&$8.57e-16$&$-6.08e-16$&$-1.13e-17$&$9.40e-10$&$-8.77e-10$&$-8.77e-10$\\
\end{tabular}
\label{tab:Tmv-B10}

\vspace{1em}

\caption{Numeric values for $T^{\hat{\mu}\hat{\nu}}$ when $B=100$:}
\begin{tabular}{l|r|r|r|r|r|r}
&$Basic$&$Basic$&$Basic$&$FRW\,Mod$&$FRW\,Mod$&$FRW\,Mod$\\$r\,(kly)$&$\rho\,(J/m^3)$&$P_{r}\,(N/m^2)$&$P_{\Omega}\,(N/m^2)$&$\rho\,(J/m^3)$&$P_{r}\,(N/m^2)$&$P_{\Omega}\,(N/m^2)$\\\hline \\3.24e-09&$9.48e+13$&$1.66e+15$&$1.46e+15$&$9.48e+13$&$1.66e+15$&$1.46e+15$\\8.42e-08&$4.07e+11$&$7.23e+12$&$2.37e+11$&$4.07e+11$&$7.23e+12$&$2.37e+11$\\2.19e-06&$9.96e+08$&$1.12e+10$&$-3.07e+07$&$9.96e+08$&$1.12e+10$&$-3.07e+07$\\5.69e-05&$2.06e+06$&$1.60e+07$&$-8.70e+04$&$2.06e+06$&$1.60e+07$&$-8.70e+04$\\1.48e-03&$3.90e+03$&$2.28e+04$&$-1.32e+02$&$3.90e+03$&$2.28e+04$&$-1.32e+02$\\3.85e-02&$7.05e+00$&$3.25e+01$&$-1.96e-01$&$7.05e+00$&$3.25e+01$&$-1.96e-01$\\1.00&$1.23e-02$&$4.62e-02$&$-2.90e-04$&$1.23e-02$&$4.62e-02$&$-2.90e-04$\\26.00&$2.10e-05$&$6.56e-05$&$-4.29e-07$&$2.10e-05$&$6.56e-05$&$-4.29e-07$\\676.00&$3.52e-08$&$9.28e-08$&$-6.34e-10$&$3.62e-08$&$9.20e-08$&$-1.51e-09$\\1.76e+04&$5.82e-11$&$1.31e-10$&$-9.38e-13$&$9.98e-10$&$-7.46e-10$&$-8.78e-10$\\4.57e+05&$9.52e-14$&$1.85e-13$&$-1.39e-15$&$9.40e-10$&$-8.77e-10$&$-8.77e-10$\\1.19e+07&$1.54e-16$&$2.60e-16$&$-2.05e-18$&$9.40e-10$&$-8.77e-10$&$-8.77e-10$\\1.60e+07&$8.57e-17$&$1.43e-16$&$-1.13e-18$&$9.40e-10$&$-8.77e-10$&$-8.77e-10$\\
\end{tabular}
\label{tab:Tmv-B100}

\end{table}

\begin{table}[h]
\caption{Numeric values for recalculated $H_0$ and $q_0$, naively using the unmodified Friedmann equations and the FRW Mod values for $\rho$ and pressures to recalculate $H_0$ and $q_0$ from input $H_0$ and $q_0$. Inputs are taken from \cite{Camarena_2020} Eqs. B1 and B2 (somewhat arbitrarily; a virtue of this source is that $H_0$ and $q_0$ are presented consecutively for ease of reference). Uncertainties are those associated with the source's calculation of $H_0$ and $q_0$ from a given astronomical dataset; the uncertainty of these parameters across the literature is much larger. The fact that the recalculated $H_0$ and $q_0$ are well within the uncertainty of the input values indicates that the FRW Mod metric has no numerically significant disagreement with the unmodified RW metric at cosmological distances.}
\begin{tabular}{l|c|r|r|r}
Parameter & Input                                                & Output             & Output           & Output               \\
               &                                                         & $B=1$             & $B=10$          &  $B=100$           \\\hline
$H_0$   & $ (8.0665 \pm .1665)e-27  \,m^-1$        &  $8.0664e-27$   &  $8.0663e-27$  &   $8.0663e-27$   \\
            & $ (74.62 \pm 1.54 km\,s^-1\,Mpc^-1) $  &  (74.62)            &     (74.62)        &  (74.62)               \\\hline
$q_0$   &           $ -0.90 \pm 0.22$                       &   $-0.899992$    &  -0.899999   &      -0.900000                    \\
\end{tabular}
\label{tab:fried}
\end{table}

\begin{table}[!htb]
\caption{Numeric values for $|T^{\hat{0}\hat{1}}|$ (radial momentum density $\times c$), which is only present in the FRW Mod version of the metric, and the numeric evaluation of the radial WEC condition $\rho + P_r - 2\left|mo_r\right|$. All evaluated $B$ have the same momentum to about 0.001\% (though the WEC condition is different because $\rho$ and $P_r$ are different).  For every evaluated $B$ and $r$, the WEC condition is positive.} 
\begin{tabular}{l|r|r|r|r}
$r\,(kly)$ & $\left|mo_r\right| \times c \,(J/m^3)$ & WEC, B=1 & WEC, B=10 & WEC, B=100\\
\hline
3.24e-09&$1.22e+06$&$3.48e+15$&$1.92e+15$&$1.76e+15$\\8.42e-08&$9.36e+02$&$1.51e+13$&$8.31e+12$&$7.63e+12$\\2.19e-06&$1.36e+00$&$2.41e+10$&$1.32e+10$&$1.22e+10$\\5.69e-05&$2.09e-03$&$3.58e+07$&$1.97e+07$&$1.81e+07$\\1.48e-03&$5.96e-06$&$5.30e+04$&$2.91e+04$&$2.67e+04$\\3.85e-02&$1.19e-07$&$7.83e+01$&$4.31e+01$&$3.96e+01$\\1.00&$4.43e-09$&$1.16e-01$&$6.37e-02$&$5.85e-02$\\26.00&$1.70e-10$&$1.71e-04$&$9.43e-05$&$8.66e-05$\\676.00&$6.54e-12$&$2.54e-07$&$1.40e-07$&$1.28e-07$\\1.76e+04&$2.52e-13$&$4.37e-10$&$2.69e-10$&$2.52e-10$\\4.57e+05&$9.68e-15$&$6.33e-11$&$6.31e-11$&$6.30e-11$\\1.19e+07&$3.72e-16$&$6.28e-11$&$6.28e-11$&$6.28e-11$\\1.60e+07&$2.76e-16$&$6.28e-11$&$6.28e-11$&$6.28e-11$\\
\end{tabular}
\label{tab:Tmv-mo-r}
\end{table}



\subsection{Particle density of the dark pressure galaxy.}\label{sec:particle-density}

This section continues numeric analysis of $T^{\hat{\mu}\hat{\nu}}$ by evaluating the particle density implied by the $\rho$ of the FRW Mod metric, to demonstrate that the density falls in a physically reasonable range when using input parameters based on the Milky Way. Note that at sub-galactic scales, the FRW Mod's $T^{\hat{\mu}\hat{\nu}}$ is almost identical to the basic $T^{\hat{\mu}\hat{\nu}}$.

Fig.~\ref{fig:protons-per-cm3} shows $\rho$ for each $B$, translated into equivalent-mass density units of protons/cm\textsuperscript{3}. Near the Sun at $r=26\,kly$, $B=1$ gives $\approx$ 14 protons/cm\textsuperscript{3}, $B=10$ gives $\approx$ 1.4 protons/cm\textsuperscript{3}, and $B=100$ gives $\approx$ 0.14 protons/cm\textsuperscript{3}. This is a reasonable range of numbers for the density of the interstellar medium (ISM) in this region\cite{Linsky_2023}\footnote{Linsky et.al., gives about 0.1 protons/cm\textsuperscript{3} for the current value, but notes that this value could have been as high as $10^4$ protons/cm\textsuperscript{3} in the past, when the Sun may have traversed colder regions of the ISM}.

\subsection{Mass contained within the dark pressure galaxy.}\label{sec:mass-contained}
These $\rho$ also describe the mass of the  central galaxy within a given radius. Integrate $\rho$ out to some radius $r'$:

\begin{equation}
M_{in}(r') = \int^{r'}_{2M}\int^{\pi}_0\int^{2\pi}_0 \rho \sqrt{-g}\times d\phi d\theta dr
\label{eq:M-in-generic}
\end{equation}
where $\sqrt{-g}$ is the `volume element' and $g$ is the determinant of the metric.

At first blush, this looks like a daunting integral. The full formula for $\rho$ of the FRW Mod metric is complicated, and the volume element adds difficultly.  However, the results in Tables \ref{tab:Tmv-B1}, \ref{tab:Tmv-B10}, \ref{tab:Tmv-B100} show that $\rho$ is largely unaffected by the FRW modification within 200,000 lightyears of the center, so we can get a good approximation of galactic mass by using the Basic result for $\rho$. Furthermore since we are using the orthonormal $\rho = T^{\hat{0}\hat{0}}$, we can use the volume element of flat space in spherical coordinates, namely $\sqrt{-g} = r^2 \sin\theta$.  The integral is then:

\begin{equation}
M_{in}(r) = \int^r_{2M}\int^{\pi}_0\int^{2\pi}_0 \left(\frac{v^2\,}{4\pi\, B\, r^2} \left(\mathrm{log}\!\left(\frac{r}{2\, M}\right) + 1\right)\right)\times r^2 \sin\theta \times d\phi\ d\theta dr 
\end{equation}

The solution is:

\begin{equation}
M_{in}(r) = r\frac{v^2 }{B} \mathrm{log}\!\left(\frac{r}{2\, M}\right) + M
\label{eq:solve-Mofr-basic}
\end{equation}

where $M$ is the arbitrary constant of integration, and corresponds to the mass of the central supermassive black hole, as before.  Note that $M_{in}$ appears directly in the metric component $g_{rr}$:

\begin{equation}
g_{rr\,basic} = \left(  1  - \frac{2M}{r} - 2\frac{v^2 }{B} \mathrm{log}\!\left(\frac{r}{2\, M}\right) \right)^{-1}  = \left( 1 - \frac{2M_{in}(r)}{r}\right)^{-1}
\label{eq:show-in-grr}
\end{equation}

Evaluating $M_{in}$ numerically at $r=$ 26 kly from galactic center yields a total mass of 2100 billion solar masses  ($M_\odot$) within 26 kly for $B=1$. Evaluation of $B=10$ yields $210$ billion $M_\odot$, a far smaller result that is (as expected) inversely proportional to $B$. $B=100$ yields 21 billion $M_\odot$. Recall that the galactic orbital speed at high $r$ is independent of $B$. Therefore, when the $T^{\hat{\mu}\hat{\nu}}$ of a spacetime contains nonzero components other than $T^{\hat{0}\hat{0}}$ ,  21 billion ($M_\odot$) may be sufficient mass to produce the flattened rotation curve shown in Fig.~\ref{fig:V_of_r}.

 For context, the stellar mass of the galaxy's inner 3 kpc (9.8 kly) has been estimated at $24$ billion $M_\odot$ \cite{McMillan_2011} (fairly close to the 21 billion ($M_\odot$) of $B=100$, but implying that  $B=100$ would be too high of a value for $B$ if we aimed to mimic this estimate precisely).  A more recent estimation from Gaia data (\cite{beordo_crosta}, Table 2), gives a non-dark-matter mass of about $15$ billion $M_\odot$ for the region between 15 and 62 kly (4.6 to 19 kpc) using both the baryonic mass estimates of a Milky Way Classical model, and mass from a model based on \cite{BALASIN_2008}.  By contrast, evaluating Eq.~\ref{eq:solve-Mofr-basic} for the same radial region and $B=100$ yields about $40$ billion $M_\odot$, a higher value.  Note, however, that the \cite{beordo_crosta} mass values are for an equatorial disc with a vertical half-width of 0.65 kly (0.2 kpc), whereas Eq.~\ref{eq:solve-Mofr-basic} applies to a spherical mass distribution with considerably larger volume.  For this reason, stellar mass density may be a better measure of realism. In \cite{mckee-2015}, stellar mass density in the vicinity of the Sun is given as $0.084 \pm 0.012 M_\odot / \texttt{pc}^{3}$, whereas the $B=100$ version of our model in the same units yields a $\rho$ of $0.034 M_\odot / \texttt{pc}^{3}$.  

If we liked, we could fine-tune $B$ to mimic  \cite{mckee-2015}'s stellar density value precisely (this requires $B = 41.1$), but this would miss this point:  The dark pressure metric is not intended to be an exact galactic model, and although it is tunable, it is \emph{not} fine-tuned:  Numeric values of $\rho$ are created with realistic Milky Way inputs and a broad, un-tuned range of $B$, and we present $\rho$ to demonstrate that, without fine-tuning, $\rho$ is compliant to an order of magnitude with realistic Milky Way densities.  Ultimately, the purpose of demonstrating broad realism is to hone the point that Milky Way-like orbits need not depend on mass alone, by pre-empting objections that would apply if $\rho$ was extremely unrealistic for the Milky Way despite being created with Milky Way inputs.

\subsection{Hydrostatic equilibrium}\label{sec:gtov}

Because the pressure is anisotropic, the dark pressure metric is not a solution to the classic Tolman-Oppenheimer-Volkoff (TOV) equation for hydrostatic equilibrium. A Generalized TOV equation (GTOV) was derived by Riazi, et al  \cite{Riazi_2016} (Eq.~\ref{eq:gtov}) to address anisotropic systems. Riazi's GTOV's derivation is not completely generalized, in that the derivation assumes a metric where $g_{tt} = -g_{rr}^{-1}$.  The dark pressure metric does not have this form, but it is nonteheless instructive to evaluate the GTOV equation for the Basic dark pressure metric. We excluded the FRW Mod dark pressure metric from the GTOV analysis, since it features cosmological expansion and is therefore not in hydrostatic equilibrium. However, since the FRW Mod's  $T^{\hat{\mu}\hat{\nu}}$ is nearly identical to the Basic  $T^{\hat{\mu}\hat{\nu}}$ at galactic scales, results for the Basic metric ought to be fairly representative of results for the FRW Mod at galactic scales, where the  FRW Mod's $T^{\hat{\mu}\hat{\nu}}$ should have local near-equilibrium.

Riazi's GTOV has the form

\begin{equation}
-\frac{dP_r}{dr} = \frac{(\rho + P_r)(M_{in}(r) + 4\pi r^3 P_r}{r^2 (1-2 M_{in}(r)/r)} + \frac{2(P_r - P_{\Omega})}{r}
\label{eq:gtov}
\end{equation}

where, as before, $\rho$ is mass-energy density, $P_r$ is radial pressure, and $M_{in}(r)$ is the mass contained within radius $r$ (including both the central black hole of mass $M$ and the mass of material outside the center, given by Eq.~\ref{eq:solve-Mofr-basic}).  $P_{\Omega}$ is the angular (tangential) pressure.  

Analytic expressions for $\rho$, $P_r$, $P_{\Omega}$, and $\frac{dP_r}{dr}$ are given in Appendix~\ref{app:full-Tmv}. $P_{\Omega}$ and $\frac{dP_r}{dr}$ are both intricate multi-line formulas.  Despite this, Eq.~\ref{eq:gtov} can be analytically evaluated with these intricate inputs and a symbolic software package.\footnote{For rigor, such analysis should begin by producing $\rho$, $P_r$, $P_{\Omega}$, and $\frac{dP_r}{dr}$ directly from the dark pressure metric, rather than attempting transcription or piecewise copying of these very long formulas.}

We find that the Basic dark pressure metric produces a stress-energy tensor which is an exact solution to the GTOV equation of  Eq.~\ref{eq:gtov}; i.e. the left and right sides are always equal.  This indicates that the Basic dark pressure stress-energy tensor is a valid representation of an imperfect fluid in hydrostatic equilibrium. This analytic result was supported by a separate numeric analysis of Eq.~\ref{eq:gtov} that used the numeric inputs of Sec.~\ref{sec:inputs}: The left and right sides were found to be equal to machine precision.

Since these $\rho$, $P_r$, $P_{\Omega}$, and $\frac{dP_r}{dr}$ represent a valid imperfect fluid, it follows that the Basic dark pressure metric solves the Einstein Field Equations for an imperfect fluid with these characteristics.  

\section{Discussion and conclusion.}\label{sec:discussion}

The metric

\begin{equation}\label{eq:metric-final}
ds^2 =  -\left( 1-\frac{2M}{r} + 2 v^2 \log\frac{r}{2M}  \right) dt^2 
+ a^2(t)\left(\frac{1}{ 1-\frac{2M}{r} - \frac{2}{B} v^2 \log\frac{r}{2M} - \kappa r^2 } dr^2
 + r^2 d\Omega^2 \right)
\end{equation}

can represent energetically realistic spacetimes which have vastly different total masses, but essentially the same orbital rotation curves.

Key features of the metric are:
\begin{itemize}

\item Mass-energy density $\rho$ of the spacetime is `tunable' by the free parameter $B$, with  $\rho \propto \frac{1}{B}$

\item The speed of stable circular orbits approaches the constant $v$ at high $r$, or $\frac{1}{a(t)} v$ if cosmological considerations are included.

\item The spacetime appears to meet the Weak Energy Condition at all points for a wide range of $B$.

\item The spacetime is not asymptotically flat, but is asymptotically indistinguishable from a Friedmann-Robertson-Walker cosmological metric. At non-cosmological $r$, the cosmological features are numerically unimportant.

\item A single analytic metric covers the entire spacetime smoothly, from the event horizon of the central black hole to the cosmological event horizon.

\item When input with standard physical parameters representative of the Milky Way galaxy, this metric produces estimates for the mass-energy density $\rho$ and contained mass $M_{in}(r)$ of the Milky Way. $\rho$ and $M_{in}(r)$ are in a reasonably realistic range, with a linear dependence on $B$.

\item When cosmological features are excluded from the metric, this spacetime is an exact solution to the GTOV equation for hydrostatic equilibrium.

\item There is pressure throughout the spacetime. 
\end{itemize}

The pressure is the primary reason this metric is most enlightening as a toy model, rather than a fully realistic galactic model.  The dark pressure does not correspond to a known physical phenomenon.

As mentioned in the Introduction, pressure is necessary to keep this metric static, because it does not rotate, so it would collapse without pressure to hold it up \cite{cooperstock}. It is possible that this metric could be altered to rotate, and that such an alteration could remove or reduce the radial pressure \cite{cooperstock} \cite{mallary-spiral}. However, initial attempts to produce rotation via the methods of Azreg-Ainou \cite{azreg-ainou} affected angular pressure more than radial pressure. Ultimately, rotational alterations are outside the scope of this work. Even without rotation, the dark pressure metric may be useful as a toy model, since general relativity problems sometimes make use of physically unlikely toy models to study rotating systems, as discussed in \cite{toy-model} and implemented in \cite{Mallary_2018}. The fact that the dark pressure metric does contain a fair amount of realism (i.e. for realistic inputs it shows WEC compliance, lack of conflict with the FRW model, and reasonable particle and stellar densities) makes it more useful as both a toy and a demonstration tool.

A more speculative interpretation is that `dark pressure' has physicality in its own right. In our galaxy, it is unlikely that pressure of this magnitude could come from the interstellar medium: Although the dark pressure is small in human terms (on the order of micro-to-milli-Pascals in the vicinity of the Sun), it is orders of magnitude higher than recent estimates of pressure in the Very Local Interstellar Medium (VLISM) \cite{Linsky_2023}, which are $23000\pm5500\,K/cm^3$ (about $3.2\times 10^{-13} Pa$).  Furthermore, dark pressure is predominantly negative for a reasonable range of $B$. This is not inherently a problem for realism. Negative pressure (tension) may represent tidal forces.  For example, consider a proton halfway in between two 1 $M_\odot$ ($2\times10^{30}$ kg) stars which are 4 lightyears apart, so that the proton is 2 lightyears from each ($r = 1.89\times10^{16}$ m). In a Newtonian approximation, the proton experiences a force toward each star of $F_{each} = -G M_\odot m_p / r^2 = -6.25 \times 10^{-40} N$, or a tension force of twice this amount ($F_{total} = -1.25\times 10^{-39} N$). Taking the mass radius of the proton to be $7.13\times10^{-16}$ m gives a cross-sectional area of $A_p = 1.60 \times 10^{-30} m^2$, so this proton is under a negative tidal pressure of $F_{total} / A_p = -0.781\times10^{-9}$ Pa.  This tidal tension is a few orders of magnitude smaller than the `dark pressure' in the vicinity of the Sun, but several orders of magnitude higher than the VLISM pressures. Given that the negative tidal tension is much higher than the positive VLISM pressure, it is worth considering that tidal tensions may play a non-negligible role in galactic dynamics: `Pressureless' galactic models are not necessarily desireable, and negative pressures may be more important than positive pressures. Nonetheless, the dark pressure metric with its high negative presssures remains a toy.

Note that these considerations of pressure also apply to the BG/Crosta galactic model (\cite{BALASIN_2008}, \cite{Crosta_2020}, \cite{beordo_crosta}), which is not pressureless in its current form. Analysis of BG/Crosta in Appendix \ref{app:crosta-pressure} shows that it cannot represent nonzero mass in the galactic plane without anisotropic pressure (including negative pressure), or a reworking of its conformal factor, which is explicity approximated as constant in the region of interest. The BG/Crosta pressure is on the same scale as the pressure of the dark pressure metric, when both metrics are set to produce the same mass-energy density: I.e. a density of 1 proton per $cm^3$ produces a $\phi$ pressure of about 50 $\mu Pa$ in the BG/Crosta model (low, but much too high for the VLISM pressure or tidal tensions on protons). 

Finally, it is possible for pervasive pressure to be difficult to detect. Carroll (\cite{carroll},  pg.~168) states that pressure can only act directly in the presence of a pressure gradient, and if this pressure is perfectly smooth then it is only detectable by its gravitational effect.  The dark pressure metric presented in this work has a pressure gradient which is smooth and extremely small outside the galactic center. However, the pressure's anisotropy presents a problem for this interpretation, since non-isotropic pressure might be detectable via a change in sensor orientation.

Regardless of whether `dark pressure' has physical reality in a Milky Way-like galaxy, or is merely an undesirable feature of a toy model, the dark pressure metric appears to be a valid GR solution, representing a spacetime configuration that could exist in our universe. This metric with its `tunable' mass and consistent orbits simply demonstrates that orbits in GR need not depend on mass alone. As a concept, this should not be surprising, since GR is fundamentally tensorial:  It is reasonable that the stress-energy tensor as a whole should contribute to orbits, not merely the $T^{00}$ mass component.  This concept underlies geometry-based galactic models like those of BG/Crosta, and is worthy of further research even if specific geometry-based models fall short of realism.  

With this work, we aim to stimulate further interest and research in geometry-based models. Slight modifications to the geometry of spacetime can have surprising gravitational effects, even in regions where the gravitational field is weak.

\section{Acknowledgements and Data Availability Statement}\label{sec:acknowledgements}
The author would like to thank the U\textsuperscript{2}GRC gravity group and attendees of the 2025 Eastern Gravity Meeting for discussions. 

All data that support the findings of this study are included within the article.

\begin{appendices}

\section{Unit conversions}\label{app:units}
Geometerized units are scaled such that $G=c=1$, and all units are expressed relative to meters. This system of units is discussed in depth by \cite{Misner1973} and \cite{natural-units}. Numeric inputs and outputs of the stress-energy are in geometerized units. Geometerized units of energy density are converted to SI units with the conversion factor:

\begin{equation}
\texttt{result in }\, \frac{J}{m^3}\, \texttt{ or } \, \frac{N}{m^2} = \frac{c^4}{G} \times \, \texttt{result in geometerized units}
\end{equation}

For the mass-energy density $\rho = T^{\hat{0}\hat{0}}$, additional conversion can be done to units of protons / $\texttt{cm}^3$:

\begin{equation}\label{eq:ppcm3-toJm3}
\rho\,\texttt{ in }\, \frac{\texttt{protons}}{cm^3}= \frac{10^{-6}}{ m_{proton}c^2} \times \, \rho \, \texttt{ in }\frac{J}{m^3}
\end{equation}
where $m_{proton} = 1.673 \times 10^{-27}$ kg.

However, note that converting  $\rho$ to protons / $\texttt{cm}^3$ is only meaningful in regions of space where we expect $\rho$ to be dominated by baryonic matter.

\section{Full analytic form of stress-energy tensor components and related values}\label{app:full-Tmv}
The full analytic form of the Basic metric's orthonormal stress-energy tensor $T^{\hat{\mu}\hat{\nu}}$ is presented here.  Also included is the analytic form of the derivative of radial pressure, to enable easier verification of the Basic metric's compliance with the Generalized TOV equation (Sec.~\ref{sec:gtov}).

For brevity, let $R = 2M$ and $L = \log\left({\frac{r}{R}}\right)$.  $T^{\hat{\mu}\hat{\nu}}$ then has the form

\begin{equation}
T^{\hat{\mu}\hat{\nu}} = 
\begin{bmatrix}
    \rho & 0 & 0 & 0 \\
    0 & P_r & 0 & 0 \\
    0 & 0 & P_{\Omega} & 0 \\
    0 & 0 & 0 & P_{\Omega}
\end{bmatrix}
\end{equation}

where

\begin{equation}
\rho = \left(\frac{L+ 1}{4 \pi B\, r^2}\right)\, v^2 
\label{eq:rho-appendix}
\end{equation}

\begin{equation}
P_r = 	 -\frac{v^2\,\left(\frac{R}{r}+\frac{2\,L\,v^2}{B}-1\right)\,\left(L\,r-B\,r+B\,R+B\,L\,R+2\,L^2\,r\,v^2+2\,L\,r\,v^2\right)}{4\,r\,\pi \,\left(2\,L\,r\,v^2-B\,r+B\,R\right)\,\left(2\,L\,r\,v^2-R+r\right)}
\label{eq:Pr-appendix}
\end{equation}

\begin{align}
P_{\Omega} = &\\ \nonumber 
 &\frac{v^2}{16\,\pi \,B\,r^2\,{\left(2\,L\,r\,v^2-R+r\right)}^2} \times  \\
&
\begin{bmatrix}
3\,R\,r-2\,r^2\,v^2+B\,R^2-L\,R^2-R^2 \\
 + 2\,L\,R\,r-2\,B\,r^2\,v^2-8\,L\,r^2\,v^2-B\,R\,r-2\,r^2-B\,L\,R^2 \\
+ 2\,R\,r\,v^2-8\,L^2\,r^2\,v^4+2\,B\,R\,r\,v^2+10\,L\,R\,r\,v^2 \\
 +4\,L^2\,R\,r\,v^2+ 4\,B\,R\,r\,L^2\,v^2+2\,B\,R\,r\,L\,v^2+2\,B\,R\,r\,L
\end{bmatrix} \nonumber 
\label{eq:Pang-appendix}
\end{align}

\begin{align}
\frac{dP_r}{dr} = &\\ \nonumber 
 &\frac{v^2}{4\,\pi \,B\,r^3\,{\left(2\,L\,r\,v^2-R+r\right)}^2} \times \\
&
\begin{bmatrix}
R\,r-2\,r^2\,v^2-B\,R^2-2\,B\,r^2+2\,L\,r^2-r^2-2\,B\,r^2\,v^2 \\
+3\,B\,R\,r-L\,R\,r+ 8\,L^3\,r^2\,v^4+4\,L^2\,r^2\,v^4+8\,L^2\,r^2\,v^2-2\,B\,L\,R^2 \\
+ 2\,R\,r\,v^2 -2\,R\,L^2\,r\,v^2-4\,B\,L\,r^2\,v^2+2\,R\,L\,r\,v^2+2\,B\,R\,r\,v^2 \\
+ 6\,B\,R\,r\,L^2\,v^2+6\,B\,R\,r\,L\,v^2+3\,B\,R\,r\,L
\end{bmatrix} \nonumber 
\end{align}

The formula for $\rho$ and $P_r$ are the same as those given in Eq.~\ref{eq:rho}\&\ref{eq:Pr}. This $T^{\hat{\mu}\hat{\nu}}$ was produced from the Basic Metric and is a solution to the Generalized TOV equation (GTOV), as can be shown with a computer algebra package. Solving the GTOV indicates that this is a valid stress energy tensor for an imperfect fluid in hydrostatic equilibrium. 

The Basic metric's analytic compliance with the WEC is discussed in Sec.~\ref{sec:results}.  The WEC requirement in the angular directions is $\rho + P_{\Omega} \geq 0$. For the Basic metric, 
\begin{equation}\label{eq:WEC-tan}
\rho + P_{\Omega} = A(BX + Y)
\end{equation}

where 

\begin{equation}\label{eq:A}
A = \frac{v^2}{16\,B\,r^2\,\pi \,{\left(2\,L\,r\,v^2-R+r\right)}^2}
\end{equation}

\begin{equation}\label{eq:X}
X = 4 L^2 R r v^2 - L R^2 + 2 L R r v^2 + 2 L R r + R^2 + 2 R r v^2 - R r - 2 r^2 v^2 \\
\end{equation}

\begin{align}\label{eq:Y}
Y = & 16 L^3 r^2 v^4 - 12 L^2 R r v^2 + 8 L^2 r^2 v^4 + 16 L^2 r^2 v^2 + 3 L R^2 - 6 L R r v^2 - 6 L R r\\ \nonumber
      &+ 8 L r^2 v^2 + 4 L r^2 + 3 R^2 + 2 R r v^2 - 5 R r - 2 r^2 v^2 + 2 r^2
\end{align}
Eq.~\ref{eq:WEC-tan} cannot be readily analyzed without choosing numeric inputs, in particular a numeric value of $B$ and a valid range for $r$. However, note that the highest-order term in $X$ is $-2 r^2 v^2$, which is negative.  This means that $BX$ is negative for sufficiently high $B$ and $r$, i.e. making $B$ too high may cause a WEC violation within the valid range of $r$.

To verify the statement in Sec.~\ref{sec:gtt} that pressures depend upon both $f(r)$ and $h(r)$ in the modified Schwarzchild metric of Eq.~\ref{eq:metric-raw}, the orthonormal radial pressure of this form of the metric is given by

\begin{align}
P_r &= \\ \nonumber
&\frac{r\,f\left(r\right)-2\,M\,h\left(r\right)+r^2\,\frac{\partial }{\partial r} h\left(r\right)+r\,h\left(r\right)\,f\left(r\right)+r^2\,f\left(r\right)\,\frac{\partial }{\partial r} h\left(r\right)-2\,M\,r\,\frac{\partial }{\partial r} h\left(r\right)}{r^2\,\left(r-2\,M+r\,h\left(r\right)\right)}
\end{align}

\section{Pressure in the `pressureless' BG/Crosta model}\label{app:crosta-pressure}

For reference, the spacetime metric used by the BG/Crosta model is:

\begin{equation}\label{eq:BG-metric}
ds^2 = -(dt - N(r,z) d\phi)^2 + r^2 d \phi^2 + e^{\nu(r,z)}(dr^2 + dz^2)
\end{equation}

As stated in the Introduction, this model is not inherently pressureless, since the Einstein equations depend on derivatives of  $\nu(r,z)$, which is never given analytic form. This appendix expands that statement.

The original BG work states that $\nu$ can be approximated as constant `for most practical purposes'. Crosta, et al. (\cite{Crosta_2020}, \cite{beordo_crosta}), acknowledge the BG/Crosta model's conformal factor $e^{\nu}$ is not constant, but approximate it as constant due to insufficient data from which to estimate it.  Additionally, \cite{beordo_crosta} concludes that the `assumption of a constant (with $r$) value for  $e^{\nu}$ holds' in the plane of the disk over a wide range of galactic-scale $r$, a conclusion reached by requiring the BG/Crosta model to fit observed rotation curves in the galactic plane. This Appendix examines the consequences of assuming constant $e^{\nu}$ .

Constant $\nu$ creates issues with the Einstein equations for a pressureless spacetime. The same Einstein equations are presented in both BG (\cite{BALASIN_2008}, Eqs. 6-10) and Crosta, et al (\cite{Crosta_2020}, Eqs. 22-26), in slightly different form.  We reproduce four of them here, using subscript derivative notation, e.g. $\partial_r N = \partial{N}/\partial{r}$. Here, $r$ is a cylindrical coordinate.

\begin{equation}\label{eq:no-rz-shear}
r \partial_z{\nu} + \partial_rN \partial_z N = 0
\end{equation}
is the requirement for no $rz$ shear,

\begin{equation}\label{eq:no-rz-pressure}
2 r \partial_r{\nu} + (\partial_r N)^2 - (\partial_z N)^2 = 0
\end{equation}
is the requirement for no radial pressure or z pressure,

\begin{equation}\label{eq:no-phi-pressure}
 \partial_{rr}{\nu} +  \partial_{zz}{\nu} +\frac{1}{2 r^2} \left( (\partial_r N)^2 + (\partial_z N)^2 \right) = 0
\end{equation}
is the requirement for no $\phi$ pressure, and

\begin{equation}\label{eq:BG-rho}
\frac{1}{r^2} \left( (\partial_r N)^2 + (\partial_z N)^2 \right) = k \rho e^{\nu}
\end{equation}
describes the mass-energy density $\rho$, and $k$ is a constant.

These formulas are greatly constrained in regions where $\nu$ is constant in $r$, because there $\partial_r{\nu} = 0$. In order for Eq.~\ref{eq:no-rz-pressure} to remain true, we need

\begin{equation}\label{eq:dnudr-is-0-constraint}
\partial_r N = \pm \partial_z N
\end{equation}
whenever $\partial_r{\nu} = 0$.

Additionally, for a galaxy model that is symmetric above and below the plane of the disc, we need the z-derivatives of $N$ and $\nu$ to be zero at $z=0$.  Together with Eq.~\ref{eq:dnudr-is-0-constraint} this means that:

\begin{equation}\label{eq:dN-is-zeroing}
\partial_r N = \pm \partial_z N = 0
\end{equation} 
whenever $\partial_r{\nu} = 0$ and $z=0$.

Taking Eq.~\ref{eq:BG-rho} to be true, this means that $\rho$ would also have to be zero in the plane of the disc in the regions where $\nu$ is constant, i.e. there is no mass in the galactic plane when pressure is zero. 

It is possible that Eq.~\ref{eq:BG-rho} contains an error: This formula appears to be based on the Ricci tensor comoponent $R_{tt}$ (\cite{BALASIN_2008}'s Eq.~A.1a) which contains no derivatives of $\nu$.  It does not seem to include the Ricci scalar term of the Einstein equation, which does depend on derivatives of $\nu$. (The other formulas from \cite{BALASIN_2008} which are included here appear correct.)  This error would be easy to miss when assuming $\nu$ is constant, since these missing derivatives of $\nu$ simply zero out. 

We recompute $\rho$ from Eq.~\ref{eq:BG-metric} via the Einstein equation.  In the orthonormal frame (as defined in Sec.~\ref{sec:math-basis}), we get:
\begin{equation}\label{eq:BG-rho-fixed}
8\pi\rho = \frac{e^{-\nu}}{4 r^2}\left( 3(\partial_r N)^2 + 3(\partial_z N)^2 - 2 r^2 \partial_{rr} \nu  - 2 r^2 \partial_{zz} \nu \right)
\end{equation}
Eq.~\ref{eq:dN-is-zeroing} remains true despite the possible error in BG/Crosta's $\rho$ equation (Eq.~\ref{eq:BG-rho}), since Eq.~\ref{eq:dN-is-zeroing} is derived from a pressureless requirement (Eq.~\ref{eq:no-rz-pressure}) and a symmetry requirement, with no dependence on $\rho$. Together with the no-$\phi$-pressure equation (Eq.~\ref{eq:no-phi-pressure}) and assumption that $\nu$ is constant in $r$, this means that: 
\begin{equation} 
\partial_{zz} \nu = 0
\end{equation}
when $z=0$, angular pressure $P_{\phi} = 0$, and $\nu$ is constant in the region. 

Applying this result to the recomputed $\rho$ equation (Eq.~\ref{eq:BG-rho-fixed}), we again find that $\rho$ must equal zero: In the BG/Crosta model, there is no mass in the galactic plane when pressure is zero and $\nu$ is constant in $r$. Since the BG/Crosta model is all about mass in the galactic plane, we cannot say that the BG/Crosta model is pressureless.

The presence of pressure in deep space is not inherently a problem for realism (see \cite{Linsky_2023} and Discussion), but this pressure should be quantified to see if it fits with observed data. To that end, we derive expressions for the pressures $P_r$, $P_z$, and $P_{\phi}$ from the metric of Eq.~\ref{eq:BG-metric} and the Einstein equations, and compare them to $\rho$. In the orthonormal frame, they are:

\begin{equation}\label{eq:BG-rho-fixed-v2}
\rho = \frac{e^{-\nu}}{32\pi r^2}\left( 3(\partial_r N)^2 + 3(\partial_z N)^2 - 2 r^2 \partial_{rr} \nu  - 2 r^2 \partial_{zz} \nu \right)
\end{equation}
\begin{equation}
P_r = -P_z = \frac{e^{-\nu}}{32 \pi r^2}\left( (\partial_r N)^2 - (\partial_z N)^2 + 2 r \partial \nu_r\right)
\end{equation}
\begin{equation}
P_{\phi} = \frac{e^{-\nu}}{32 \pi r^2}\left( (\partial_r N)^2 + (\partial_z N)^2 + 2 r^2 \partial_{rr} \nu  + 2 r^2 \partial_{zz} \nu \right)
\end{equation}

These pressures are not isotropic, and the $z$ pressure is negative when $r$ pressure is positive. Furthermore, in regions where $\nu$ is approximately constant, the derivatives of $\nu$ drop, and
\begin{equation}\label{eq:bg-crosta-pressure-phi-vs-rho}
P_{\phi} \approx \frac{1}{3}\rho
\end{equation}

This is not an insignificant amount of pressure. For a $\rho$ equivalent to 1 proton per $cm^3$ (a rough ISM density in the vicinity of the Sun), we can use Eq.~\ref{eq:ppcm3-toJm3} on Eq.~\ref{eq:bg-crosta-pressure-phi-vs-rho} to obtain 

\begin{equation}\label{eq:bg-crosta-pressure-1ppcm3}
P_{\phi} \approx 50 \, \mu Pa
\end{equation}
for the BG/Crosta metric at  $\rho\, \approx$ 1 proton per $cm^3$, when $\nu\, \approx$ constant, with $1\,Pa = 1\,J/m^3$.  

We compare this to the dark pressure metric (an explicitly pressurized spacetime). The dark pressure metric yields 1 proton per $cm^3$ in the vicinity of the Sun when we set $B=14$. For this value of $B$, setting $r$ to the Sun's radius, we get:

\begin{eqnarray}\label{eq:dp-pressure-1ppcm3}
P_{r} & \approx & -58 \, \mu Pa \\
P_{\Omega} & \approx & -3 \, \mu Pa \\
\end{eqnarray}
for the dark pressure metric at  $\rho\, \approx$ 1 proton per $cm^3$, when $B=14$, $r = 26\,kly$, with $1\,Pa = 1\,J/m^3$.  

It is interesting that the BG/Crosta metric and the dark pressure metric yield similar magnitudes of pressure for similar mass densities (at least for these locations in the spacetimes). However, as mentioned in the Discussion, these pressures are many orders of magnitude higher than measured local ISM pressures. The BG/Crosta model is not approximately pressureless: It is an imperfect fluid rather than a dust. If Crosta, et al. strive for realism, they must either refine their model or explain why these pressures are not a concern.

Hypothetically, a theoretical explanation for the BG/Crosta pressure would also apply to the dark pressure metric. Since we are not aware of any accepted explanation, we must consider the dark pressure metric a `toy model', and consider the BG/Crosta model to be a toy as well. Note that the original BG paper authors \cite{BALASIN_2008} also describe their model as a toy. The strong claims of realism made by Crosta, et al., were not made by the creators of the model.

\end{appendices}

\bibliographystyle{apalike} 
\bibliography{DarkPressureBibliography} 

@article{mckee-2015,
  title={Stars, gas, and dark matter in the solar neighborhood},
  author={Christopher F. McKee and Antonio Parravano and David J. Hollenbach},
  journal={The Astrophysical Journal},
  year={2015},
  volume={814},
  url={https://api.semanticscholar.org/CorpusID:54224451}
}

@book{	carroll,
	author={Carroll, S. M.},
	title={Spacetime and Geometry: An Introduction to General Relativity},
	year={2004},
 	publisher={Cambridge University Press},
	address={Cambridge}
}

@article{Camarena_2020,
   title={Local determination of the Hubble constant and the deceleration parameter},
   volume={2},
   ISSN={2643-1564},
   url={http://dx.doi.org/10.1103/PhysRevResearch.2.013028},
   DOI={10.1103/physrevresearch.2.013028},
   number={1},
   journal={Physical Review Research},
   publisher={American Physical Society (APS)},
   author={Camarena, David and Marra, Valerio},
   year={2020},
   month=jan 
}

@article{cooperstock,
   title={Galactic Dynamics via General Relativity: A Compilation and New Developments},
   volume={22},
   url={https://arxiv.org/abs/astro-ph/0610370},
   DOI={10.1142/s0217751x0703666x},
   number={13},
   journal={International Journal of Modern Physics A},
   publisher={World Scientific Pub Co Pte Lt},
   author={Cooperstock, F. I. and Tieu, S.},
   year={2007},
   month=may
}

@article{planck,
   title={Planck 2018 results: VI. Cosmological parameters},
   volume={641},
   ISSN={1432-0746},
   url={http://dx.doi.org/10.1051/0004-6361/201833910},
   DOI={10.1051/0004-6361/201833910},
   journal={Astronomy and Astrophysics},
   publisher={EDP Sciences},
   author={Aghanim and etal},
   year={2020},
   month=sep, pages={A6} 
}

@conference{toy-model,
	author = {Johnson-McDaniel, N. K.},
	title = {Charged black holes in GR and beyond},
	url = {https://benasque.org/2018relativity/talks_contr/049_McDaniel.pdf},
	organization = {Benasque meeting: NR beyond GR},	
	date = {2018-06-04},
	urldate = {2025-04-12},
	year={2018},
	number={LIGO-G1801141-v1}
}

@article{Mallary_2018,
   title={Physical objects approaching the Cauchy horizon of a rapidly rotating Kerr black hole},
   volume={98},
   ISSN={2470-0029},
   url={http://dx.doi.org/10.1103/PhysRevD.98.104024},
   DOI={10.1103/physrevd.98.104024},
   number={10},
   journal={Physical Review D},
   publisher={American Physical Society (APS)},
   author={Mallary, C. and Khanna, G. and Burko, L. M.},
   year={2018},
   month=nov 
}

@book{Misner1973,
  author = {{Misner}, C. W. and {Thorne}, K. S. and {Wheeler}, J. A.},
  title = {Gravitation},
  publisher={Freeman},
  address={San Francisco},
  year = {1973}
}

@online{natural-units,
	author = {Myers, A.L.},
	title = {Natural System of Units in General Relativity.},
	url = {https://www.seas.upenn.edu/%7Eamyers/NaturalUnits.pdf},
	organization = {University of Pennsylvania},	
	date = {2016},
	urldate = {2025-04-12},
	year={2016}
}

@article{MOND,
       author = {{Milgrom}, M.},
        title = {A modification of the Newtonian dynamics as a possible alternative to the hidden mass hypothesis.},
      journal = {The Astrophysical Jouranal},
         year = 1983,
        month = jul,
       volume = {270},
        pages = {365-370},
          doi = {10.1086/161130},
       adsurl = {https://ui.adsabs.harvard.edu/abs/1983ApJ...270..365M}
}

@Inbook{Curiel2017,

author="Curiel, Erik",

editor="Lehmkuhl, Dennis
and Schiemann, Gregor
and Scholz, Erhard",

title="A Primer on Energy Conditions",

bookTitle="Towards a Theory of Spacetime Theories",

year="2017",

publisher="Springer New York",
address="New York, NY",

pages="43--104",

isbn="978-1-4939-3210-8",
doi="10.1007/978-1-4939-3210-8_3",

url="https://doi.org/10.1007/978-1-4939-3210-8_3"

}

@article{Linsky_2023,
doi = {10.3847/1538-4357/aca676},
url = {https://dx.doi.org/10.3847/1538-4357/aca676},
year = {2023},
month = {jan},
publisher = {The American Astronomical Society},
volume = {942},
number = {1},
pages = {45},
author = {Linsky, Jeffrey L. and Moebius, Eberhard},
title = {Are the Heliosphere, Very Local Interstellar Medium, and Local Cavity in Pressure Balance with Galactic Gravity?*},
journal = {The Astrophysical Journal}
}

@article{McMillan_2011,
   title={Mass models of the Milky Way: Mass models of the Milky Way},
   volume={414},
   ISSN={0035-8711},
   url={http://dx.doi.org/10.1111/j.1365-2966.2011.18564.x},
   DOI={10.1111/j.1365-2966.2011.18564.x},
   number={3},
   journal={Monthly Notices of the Royal Astronomical Society},
   publisher={Oxford University Press (OUP)},
   author={McMillan, Paul J.},
   year={2011},
   month=apr, pages={2446–2457} }

@article{Crosta_2020,
   title={On testing CDM and geometry-driven Milky Way rotation curve models with Gaia DR2},
   volume={496},
   ISSN={1365-2966},
   url={http://dx.doi.org/10.1093/mnras/staa1511},
   DOI={10.1093/mnras/staa1511},
   number={2},
   journal={Monthly Notices of the Royal Astronomical Society},
   publisher={Oxford University Press (OUP)},
   author={Crosta, Mariateresa and Giammaria, Marco and Lattanzi, Mario G and Poggio, Eloisa},
   year={2020},
   month=jun, pages={2107–2122} }

@article{BALASIN_2008,
   title={Non-Newtonian Behavior in Weak Field General Relativity for Extended Rotating Sources},
   volume={17},
   ISSN={1793-6594},
   url={http://dx.doi.org/10.1142/S0218271808012140},
   DOI={10.1142/s0218271808012140},
   number={03n04},
   journal={International Journal of Modern Physics D},
   publisher={World Scientific Pub Co Pte Ltd},
   author={Balasin, H. and Grumiller, D.},
   year={2008},
   month=mar, pages={475–488} }

@article{beordo_crosta,
    author = {Beordo, William and Crosta, Mariateresa and Lattanzi, Mario G and Re Fiorentin, Paola and Spagna, Alessandro},
    title = {Geometry-driven and dark-matter-sustained Milky Way rotation curves with Gaia DR3},
    journal = {Monthly Notices of the Royal Astronomical Society},
    volume = {529},
    number = {4},
    pages = {4681-4698},
    year = {2024},
    month = {03},
    issn = {0035-8711},
    doi = {10.1093/mnras/stae855},
    url = {https://doi.org/10.1093/mnras/stae855},
    eprint = {https://academic.oup.com/mnras/article-pdf/529/4/4681/57147041/stae855.pdf},
}

@article{Costa_2023,
   title={Reference frames in general relativity and the galactic rotation curves},
   volume={108},
   ISSN={2470-0029},
   url={http://dx.doi.org/10.1103/PhysRevD.108.044056},
   DOI={10.1103/physrevd.108.044056},
   number={4},
   journal={Physical Review D},
   publisher={American Physical Society (APS)},
   author={Costa, L. Filipe O. and Natário, José and Frutos-Alfaro, F. and Soffel, Michael},
   year={2023},
   month=aug }

@conference{mallary-spiral,
	author = {Mallary, C.},
	title = {Analytic metric for a spiraling halo},
	url = {https://web.uri.edu/gravity/egm23-schedule/},
	organization = {Eastern Gravity Meeting},	
	date = {2023-06-09},
	urldate = {2025-04-16},
	year={2023}
}

@article{Riazi_2016,
   title={A new class of anisotropic solutions of the generalized TOV equation},
   volume={94},
   ISSN={1208-6045},
   url={http://dx.doi.org/10.1139/cjp-2016-0365},
   DOI={10.1139/cjp-2016-0365},
   number={10},
   journal={Canadian Journal of Physics},
   publisher={Canadian Science Publishing},
   author={Riazi, Nematollah and Hashemi, S. Sedigheh and Sajadi, S. Naseh and Assyyaee, Shahrokh},
   year={2016},
   month=oct, pages={1093–1101} }

@misc{vogt2005,
      title={Presence of exotic matter in the Cooperstock and Tieu galaxy model}, 
      author={D. Vogt and P. S. Letelier},
      year={2005},
      eprint={astro-ph/0510750},
      archivePrefix={arXiv},
      primaryClass={astro-ph},
      url={https://arxiv.org/abs/astro-ph/0510750}, 
}

@article{azreg-ainou,
title = {Regular and conformal regular cores for static and rotating solutions},
journal = {Physics Letters B},
volume = {730},
pages = {95-98},
year = {2014},
issn = {0370-2693},
doi = {https://doi.org/10.1016/j.physletb.2014.01.041},
url = {https://www.sciencedirect.com/science/article/pii/S037026931400063X},
author = {Mustapha Azreg-Aïnou}
}

\end{document}